\documentclass[journal,10pt]{IEEEtran}
\usepackage{amsmath}
\usepackage{graphicx}
\usepackage{amsfonts}
\usepackage{caption}
\usepackage{color}
\usepackage{cuted}
\usepackage{multicol}
\usepackage{algorithmic}
\usepackage{algorithm}
\usepackage{float}
\usepackage{comment}
\usepackage{amssymb}
\usepackage{stfloats}
\usepackage{bbm}
\usepackage[caption=false]{subfig}
\usepackage{cleveref}
\usepackage[english]{babel}
\usepackage{amsthm}
\usepackage{bigints}

\theoremstyle{plain}

\theoremstyle{definition}

\theoremstyle{remark}

\usepackage{pifont}

\graphicspath{{Figures/}{../Figures/}}
\usepackage[english]{babel}
\newtheorem{theorem}{\bf Theorem}
\usepackage{enumitem}
\usepackage{parskip}
\usepackage{xcolor}
\newlist{Properties}{enumerate}{2}
\usepackage{stix}

\begin{document}
\title{A 
Gaussian-Sinc Pulse Shaping Filter for Zak-OTFS}
\author{Arpan Das, Fathima Jesbin, and Ananthanarayanan Chockalingam\thanks{The authors are with the Department of ECE, Indian Institute of Science, Bangalore 560012, India. Email: \{arpandas1,fathimaj,achockal\}@iisc.ac.in.}
\vspace{-2mm}
}
\maketitle
\begin{abstract} 
The choice of delay-Doppler domain (DD) pulse shaping filter plays an important role in determining the performance of Zak-OTFS. Sinc filter has good main lobe characteristics (with nulls at information grid points) which is good for equalization/detection, but has high side lobes which are detrimental for input-output (I/O) relation estimation. Whereas, Gaussian filter is highly localized with very low side lobes which is good for I/O relation estimation, but has poor main lobe characteristics which is not good for equalization/detection. In this paper, we propose a new filter, termed as {\em Gaussian-sinc (GS) filter}, which inherits the complementary strengths of both Gaussian and sinc filters. The proposed filter does not incur time or bandwidth expansion. We derive closed-form expressions for the I/O relation and noise covariance of Zak-OTFS with the proposed GS filter. We evaluate the Zak-OTFS performance for different pulse shaping filters with I/O relation estimated using exclusive and embedded pilots. Our results show that the proposed GS filter achieves better bit error rate (BER) performance compared to other filters reported in the literature. For example, with model-free I/O relation estimation using embedded pilot and 8-QAM, the proposed GS filter achieves an SNR gain of about 4 dB at $10^{-2}$ uncoded BER compared to Gaussian and sinc filters, and the SNR gain becomes more than 6 dB  at a coded BER of $10^{-4}$ with rate-1/2 coding.
\end{abstract}
\vspace{-2.5mm}
\begin{IEEEkeywords}
Zak-OTFS modulation, delay-Doppler domain, pulse shaping filter, noise covariance, I/O relation estimation, equalization/detection. 
\end{IEEEkeywords}
\vspace{-2mm}
\section{Introduction}
\label{sec1}
Orthogonal time frequency space (OTFS) modulation is a delay-Doppler (DD) domain modulation suited for doubly-selective channels. In multicarrier OTFS (MC-OTFS) modulation introduced in \cite{otfs1}, the information symbols in the DD domain are converted to time-frequency (TF) domain following which conversion to time domain is carried out using a legacy multicarrier modulation scheme \cite{otfs2}-\cite{h_b_mishra}. In Zak transform based OTFS (Zak-OTFS) modulation, the information symbols multiplexed in the DD domain are converted to time domain for transmission using inverse Zak transform \cite{zak_otfs1},\cite{zak_otfs2},\cite{zak_otfs3}. At the receiver, the received time domain signal is converted back to DD domain using Zak transform for data detection. In this paper, we consider Zak-OTFS. Two key aspects are central to Zak-OTFS. First, it provides a formal mathematical framework using Zak theory for describing OTFS and studying its fundamental properties \cite{zak_otfs1},\cite{zak_otfs3}. This is analogous to how Fourier theory provides an appropriate mathematical framework for describing and understanding OFDM. Second, Zak-OTFS waveform is more robust to a larger range of delay and Doppler spreads of the channel. This is because the input-output (I/O) relation in Zak-OTFS is non-fading and predictable, even in the presence of significant delay and Doppler spreads, and, as a consequence, the channel can be efficiently acquired and equalized \cite{zak_otfs2}. Recent works on Zak-OTFS have been reported in \cite{zak_otfs4}-\cite{zak_otfs8}.

An important building block in the Zak-OTFS transmitter is the DD domain transmit pulse shaping filter. The basic information-bearing carrier in Zak-OTFS is a pulse in the DD domain which is a quasi-periodic localized function. The Zak-OTFS performance is influenced by how well these pulses are localized in the DD domain. A DD filter matched to the transmit filter is used at the receiver. The `effective' channel in Zak-OTFS includes the cascade of the transmit DD filter, the physical channel, and the receive DD filter. Consequently, the choice of the pulse shaping filter influences the DD spread of the effective channel. Estimating the DD domain input-output (I/O) relation in Zak-OTFS amounts to estimating the coefficients of the effective channel. The estimated I/O relation is used for subsequent equalization/detection in the DD domain. Therefore, the pulse shape influences the performance of the two important receiver functions, namely, I/O relation estimation and equalization/detection. 
    
In the Zak-OTFS literature, the following DD pulse shaping filters have been considered: 1) sinc filter
\cite{zak_otfs2}-\cite{zak_otfs5},\cite{zak_otfs9},\cite{zak_otfs8},
2) root raised cosine (RRC) filter \cite{zak_otfs2}-\cite{zak_otfs9}, and 3) Gaussian filter \cite{zak_otfs7}. The sinc filter has the benefit of good main lobe characteristics with nulls at the Nyquist sampling points in the DD domain (i.e., nulls at the information grid points). This attribute has a positive influence on achieving good equalization/detection performance. However, sinc filter has the drawback of high side lobes which plays a negative role in I/O relation estimation. Specifically, pulse shaping filters cause aliasing between the received pilot and its own quasi-periodic replicas (a.k.a. self-interaction). Because of this, the high side lobes in the sinc filter result in increased aliasing (self-interference due to quasi-periodic replicas) that leads to poor I/O relation estimation. The RRC filter achieves reduced side lobe levels compared to sinc filter, but this side lobe reduction is achieved with bandwidth and time expansion. More the bandwidth and time expansion, better will be the side lobe reduction. The Gaussian filter, on the other hand, has the advantage of good DD localization with very low side lobe levels, but it has poor main lobe characteristics without nulls at the Nyquist sampling points. This makes the Gaussian filter superior for I/O relation estimation but inferior for equalization/detection compared to sinc and RRC filters. 

Based on the above observations, in this paper, we propose a new pulse shaping filter, termed {\em Gaussian-sinc (GS) filter}, which inherits the complementary strengths of Gaussian and sinc filters. Unlike RRC filter, the proposed filter does not incur time and bandwidth expansion. We derive closed-form expressions for the I/O relation and noise covariance of Zak-OTFS with the proposed GS filter. We evaluate the Zak-OTFS performance for different pulse shaping filters with I/O relation estimated using exclusive and embedded pilots. We consider ITU Vehicular-A (Veh-A) channel model \cite{ITU_VehA} with fractional delays and Dopplers in performance evaluation. Our simulation results show that the proposed GS filter achieves better bit error rate (BER) performance compared to other filters reported in the literature.  
For example, with model-free I/O relation estimation using embedded pilot and 8-QAM, the proposed GS filter achieves an SNR gain of about 4 dB at $10^{-2}$ uncoded BER compared to Gaussian and sinc filters, and the SNR gain becomes more than 6 dB  at a coded BER of $10^{-4}$ with rate-1/2 coding.

The rest of the paper is organized as follows. The Zak-OTFS system model and the sinc, RRC, and Gaussian filters are introduced in Sec. \ref{sec2}. The model-free I/O relation estimation using exclusive and embedded pilot frames is presented in Sec. \ref{sec3}. The proposed GS filter and the derivation of closed-form expressions for the I/O relation and noise covariance are presented in Sec. \ref{sec4}. Performance results and discussions are presented in Sec. \ref{sec5}. Conclusions and future work are presented in Sec. \ref{sec6}. 
\vspace{-1mm}
\section{Zak-OTFS System Model}
\label{sec2}
\begin{figure*}
\centering    \includegraphics[width=0.95\linewidth]{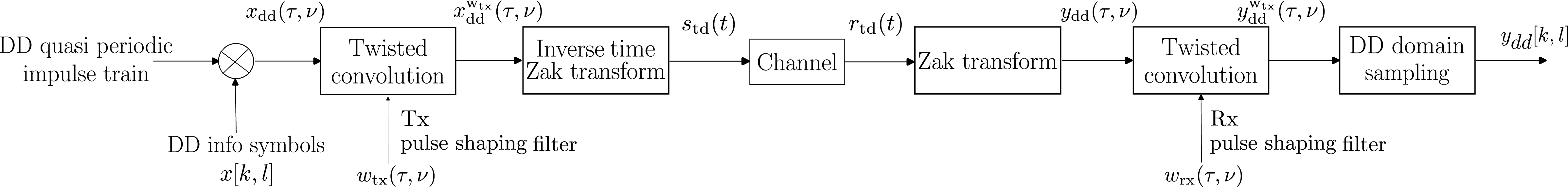}
\caption{Block diagram of Zak-OTFS transceiver.}
\label{fig1}      
\vspace{-4mm}
\end{figure*}

Figure \ref{fig1} shows the block diagram of a Zak-OTFS transceiver.
In Zak-OTFS, a pulse in the DD domain is the basic information carrier. A DD pulse is a quasi-periodic localized function defined by a delay period $\tau_{\mathrm{p}}$ and a Doppler period $\nu_{\mathrm{p}}=\frac{1}{\tau_{\mathrm{p}}}$. The fundamental period in the DD domain is defined as 
$\mathcal{D}_{0}= \{(\tau,\nu): 0\leq\tau<\tau_{\mathrm p}, 0\leq\nu<\nu_{\mathrm p}\}$,
where $\tau$ and $\nu$ represent the delay and Doppler variables, respectively. The fundamental period is discretized into $M$ bins on the delay axis and $N$ bins on the Doppler axis, as 
$\big\{(k\frac{\tau_{{\mathrm p}}}{M},l\frac{\nu_{{\mathrm p}}}{N}) | k=0,\ldots,M-1,l=0,\ldots,N-1\big\}$. The time domain Zak-OTFS frame is limited to a time duration $T=N\tau_{\mathrm p}$ and a bandwidth $B=M\nu_{\mathrm p}$. In each frame, $MN$ information symbols drawn from a modulation alphabet ${\mathbb A}$, $x[k,l]\in {\mathbb A}$, $k=0,\ldots,M-1$, $l=0,\ldots,N-1$, are multiplexed in the DD domain. The information symbol $x[k,l]$ is carried by DD domain pulse $x_{\mathrm{dd}}[k,l]$, which is a quasi-periodic function with period $M$ along the delay axis and period $N$ along the Doppler axis, i.e., for any $n,m\in\mathbb{Z}$,  
\begin{equation}
x_{\mathrm{dd}}[k+nM,l+mN]=x[k,l]e^{j2\pi n\frac{l}{N}}.
\end{equation}
These discrete DD domain signals $x_{\mathrm{dd}}[k,l]$s are supported on the information lattice 
$\Lambda_{\mathrm{dd}}=
\big\{\big(k\frac{\tau_{\mathrm p}}{M},l\frac{\nu_{\mathrm p}}{N}\big) | k,l\in \mathbb{Z}\big\}$.
The continuous DD domain information signal is given by
\vspace{-1mm}
\begin{equation}
x_{\mathrm{dd}}(\tau,\nu)=\sum_{k,l\in \mathbb{Z}} x_{\mathrm{dd}}[k,l] \delta\Big(\tau-\frac{k\tau_{\mathrm p}}{M}\Big)\delta\Big(\nu-\frac{l\nu_{\mathrm p}}{N}\Big),
\end{equation}
where $\delta(.)$ denotes the Dirac-delta impulse function. For any $n,m\in \mathbb{Z}$, we have
$x_{\mathrm{dd}}(\tau+n\tau_{\mathrm{p}},\nu+m\nu_{\mathrm{p}})=e^{j2\pi n\nu \tau_{\mathrm{p}}}x_{\mathrm{dd}}(\tau,\nu)$,
so that $x_{\mathrm{dd}}(\tau,\nu)$ is periodic with period $\nu_{\mathrm p}$ along the Doppler axis and quasi-periodic with period $\tau_{\mathrm p}$ along the delay axis.

The DD domain transmit signal $x_{\mathrm{dd}}^{w_{\mathrm{tx}}}(\tau,\nu)$ is given by the twisted convolution of the transmit pulse shaping filter $w_{\mathrm{tx}}(\tau,\nu)$ with $x_{\mathrm{dd}}(\tau,\nu)$ as $x_{\mathrm{dd}}^{w_{\mathrm{tx}}}(\tau,\nu) = w_{\mathrm{tx}}(\tau,\nu)*_{\sigma}x_{\mathrm{dd}}(\tau,\nu)$,
where $*_{\sigma}$ denotes the twisted convolution\footnote{Twisted convolution of two DD functions $a(\tau,\nu)$ and $b(\tau,\nu)$ is defined as 
$a(\tau,\nu) \ast_\sigma b(\tau,\nu) \overset{\Delta}{=} \iint a(\tau', \nu') b(\tau-\tau',\nu-\nu')e^{j2\pi\nu'(\tau-\tau')}d\tau'  d\nu'$.}. The transmitted time domain (TD) signal $s_{\mathrm{td}}(t)$ is the TD realization of $x_{\mathrm{dd}}^{w_{\mathrm{tx}}}(\tau,\nu)$, given by
$s_{\mathrm{td}}(t)=Z_{t}^{-1}\left(x_{\mathrm{dd}}^{w_{\mathrm{tx}}}(\tau,\nu)\right)$, where $Z_{t}^{-1}$ denotes the inverse time-Zak transform operation\footnote{Inverse time-Zak transform of a DD function $a(\tau,\nu)$ is defined as $Z_{t}^{-1}(a(\tau,\nu)) \overset{\Delta}{=} \sqrt{\tau_{\mathrm p}} \int_0^{\nu_{\mathrm p}} a(t,\nu) d\nu$.}. The transmit pulse shaping filter $w_{\mathrm{tx}}(\tau,\nu)$ 
limits the time and bandwidth of the transmitted signal $s_{\mathrm{td}}(t)$. The transmit signal $s_{\mathrm{td}}(t)$ passes through a doubly-selective channel to give the output signal $r_{\mathrm{td}}(t)$. The DD domain impulse response of the physical channel $h_{\mathrm{phy}}(\tau,\nu)$ is given by
\begin{equation}
h_{\mathrm{phy}}(\tau,\nu)=\sum_{i=1}^{P}h_{i}\delta(\tau-\tau_{i})\delta(\nu-\nu_{i}),
\end{equation}
where $P$ denotes the number of DD paths, and the $i$th path has gain $h_{i}$, delay shift $\tau_{i}$, and Doppler shift $\nu_{i}$. 

The received TD signal $y(t)$ at the receiver is given by $y(t)=r_{\mathrm{td}}(t)+n(t)$,
where $n(t)$ is AWGN with variance $N_{0}$, i.e., $\mathbb{E}[n(t)n(t+t')]=N_{0}\delta(t')$. The TD signal $y(t)$ is converted to the corresponding DD domain signal $y_{\mathrm{dd}}(\tau,\nu)$ by applying Zak transform\footnote{Zak transform of a continuous TD signal $a(t)$ is defined as
$Z_t\left(a(t)\right) \overset{\Delta}{=} \sqrt{\tau_p} \sum_{k \in \mathbb{Z}} a(\tau + k \tau_{\mathrm p}) e^{-j2\pi\nu k\tau_{\mathrm p}}$.}, i.e.,
\begin{eqnarray}
\hspace{-6mm}
y_{\mathrm{dd}}(\tau,\nu) = Z_{t}(y(t)) 
= r_{\mathrm{dd}}(\tau,\nu)+n_{\mathrm{dd}}(\tau,\nu),
\end{eqnarray}
where $r_{\mathrm{dd}}(\tau,\nu)=h_{\mathrm{phy}}(\tau,\nu)*_{\sigma}w_{\mathrm{tx}}(\tau,\nu)*_{\sigma}x_{\mathrm{dd}}(\tau,\nu)$ is the Zak transform of $r_{\mathrm{td}}(t)$, given by the twisted convolution cascade of $x_{\mathrm{dd}}(\tau,\nu)$, $w_{\mathrm{tx}}(\tau,\nu)$, and $h_{\mathrm{phy}}(\tau,\nu)$,  and $n_{\mathrm{dd}}(\tau,\nu)$ is the Zak transform of $n(t)$. The receiver filter $w_{\mathrm{rx}}(\tau,\nu)$ acts on $y_{\mathrm{dd}}(\tau,\nu)$ through twisted convolution to give the output 
\begin{eqnarray}
\hspace{-4mm} 
y_{\mathrm{dd}}^{w_{\mathrm{rx}}}(\tau,\nu) & \hspace{-2mm} = & \hspace{-2mm} w_{\mathrm{rx}}(\tau,\nu)*_{\sigma}y_{\mathrm{dd}}(\tau,\nu) \nonumber \\ 
& \hspace{-22mm} = & \hspace{-12mm} \underbrace{w_{\mathrm{rx}}(\tau,\nu)*_{\sigma}h_{\mathrm{phy}}(\tau,\nu)*_{\sigma}w_{\mathrm{tx}}(\tau,\nu)}_{\overset{\Delta}{=} \ h_{\mathrm{eff}}(\tau,\nu)}*_{\sigma}x_{\mathrm{dd}}(\tau,\nu) \nonumber \\ 
&\hspace{-22mm} & \hspace{-12mm} + \ \underbrace{w_{\mathrm{rx}}(\tau,\nu)*_{\sigma}n_{\mathrm{dd}}(\tau,\nu)}_{\overset{\Delta}{=} \ n_{\mathrm{dd}}^{w_{\mathrm{rx}}}(\tau,\nu)}, 
\label{cont1}
\end{eqnarray}
where $h_{\mathrm{eff}}(\tau,\nu)$ denotes the effective channel consisting of the twisted convolution cascade of $w_{\mathrm{tx}}(\tau,\nu),\ h_{\mathrm{phy}}(\tau,\nu)$, and $w_{\mathrm{rx}}(\tau,\nu)$, and $n_{\mathrm{dd}}^{w_{\mathrm{rx}}}(\tau,\nu)$ denotes the noise filtered through the Rx filter. The DD signal $y_{\mathrm{dd}}^{w_{\mathrm{rx}}}(\tau,\nu)$ is sampled on the information lattice, resulting in the discrete quasi-periodic DD domain received signal $y_{\mathrm{dd}}[k,l]$ as
\vspace{0mm}
\begin{equation}
y_{\mathrm{dd}}[k,l]=y_{\mathrm{dd}}^{w_{\mathrm{rx}}}\left(\tau=\frac{k\tau_{\mathrm p}}{M},\nu=\frac{l\nu_{\mathrm p}}{N}\right), \ \ k,l\in\mathbb{Z},
\end{equation} 
which is given by
$y_{\mathrm{dd}}[k,l]=h_{\mathrm{eff}}[k,l]*_{\sigma\text{d}}x_{\mathrm{dd}}[k,l]+n_{\mathrm{dd}}[k,l]$,
where $*_{\sigma\text{d}}$ is twisted convolution in discrete DD domain, i.e., 
$h_{\mathrm{eff}}[k,l]*_{\sigma\text{d}}x_{\mathrm{dd}}[k,l] = \sum_{k',l'\in\mathbb{Z}}h_{\mathrm{eff}}[k-k',l-l']x_{\mathrm{dd}}[k',l'] e^{j2\pi\frac{k'(l-l')}{MN}}$, where the effective channel filter $h_{\mathrm{eff}}[k,l]$ and filtered noise samples $n_{\mathrm{dd}}[k,l]$ are given by
\begin{align}
h_{\text{eff}}[k,l]=h_{\text{eff}}\left(\tau=\frac{k\tau_{p}}{M},\nu=\frac{l\nu_{p}}{N}\right), \label{discr2} \\ 
n_{\text{dd}}[k,l]=n_{\text{dd}}^{w_{\mathrm{rx}}}\left(\tau=\frac{k\tau_{p}}{M},\nu=\frac{l\nu_{p}}{N}\right).
\label{discr3}
\end{align}
Owing to the quasi-periodicity in the DD domain, it is sufficient to consider the received samples $y_{\mathrm{dd}}[k,l]$ within the fundamental period $\mathcal{D}_0$. Writing the $y_{\mathrm{dd}}[k,l]$ samples as a vector, the received signal model can be written in matrix-vector form as \cite{zak_otfs1},\cite{zak_otfs2}
\begin{equation}
\mathbf{y}=\mathbf{H_\text{eff}x}+\mathbf{n},
\label{sys_mod}
\end{equation}
where $\mathbf{x,y,n} \in\mathbb{C}^{MN\times 1}$, such that their $(kN+l+1)$th entries are given by $x_{kN+l+1}=x_{\mathrm{dd}}[k,l]$, $y_{kN+l+1}=y_{\mathrm{dd}}[k,l]$, $n_{kN+l+1}=n_{\mathrm{dd}}[k,l]$, and $\mathbf{H}_{\text{eff}}\in\mathbb{C}^{MN\times MN}$ is the effective channel matrix such that
\vspace{-1mm}
\begin{eqnarray}
\mathbf{H}_\text{eff}[k'N+l'+1,kN+l+1] & \hspace{-2mm} = & \hspace{-2mm} \sum_{m,n\in\mathbb{Z}}h_{\mathrm{eff}}[k'-k-nM, \nonumber \\
& \hspace{-45mm} & \hspace{-35mm} l'-l-mN]e^{j2\pi nl/N}e^{j2\pi\frac{(l'-l-mN)(k+nM)}{MN}},
\label{eqn_channel_matrix}
\vspace{-4mm}
\end{eqnarray}
where $k',k=0,\ldots,M-1$, $l',l=0,\ldots,N-1$. 

\vspace{-2mm}
\subsection{DD pulse shaping filters}
In the absence of pulse shaping, i.e., $w_\text{tx}(\tau,\nu)=\delta(\tau,\nu)$, the transmit signal has infinite time duration and bandwidth. Pulse shaping limits the time and bandwidth of transmission. We consider transmit DD pulse shaping filters of the form $w_\text{tx}(\tau,\nu)=w_1(\tau)w_2(\nu)$ \cite{zak_otfs2},\cite{zak_otfs6}. The time duration $T'$ of each frame is approximately related to the spread of $w_2({\nu})$ along the Doppler axis as $\frac{1}{T'}$. Likewise, the bandwidth $B'$ is approximately related to the spread of $w_1(\tau)$ along the delay axis as $\frac{1}{B'}$. That is, a larger bandwidth and time duration implies a smaller DD spread of $w_\text{tx}(\tau,\nu)$, and hence a smaller contribution to the spread of $h_\text{eff}(\tau,\nu)$. Sinc, RRC, and Gaussian filters have been considered in the Zak-OTFS literature and are described below. 

{\em Sinc filter:}
For sinc filter, $w_1({\tau})$ and $w_2({\nu})$ are given by
$w_1({\tau}) = \sqrt{B}\text{sinc}(B\tau)$ and $w_2({\nu}) = \sqrt{T}\text{sinc}(T\nu)$,
so that 
\begin{equation}
w_\text{tx}(\tau,\nu)
=\underbrace{\sqrt{B}\text{sinc}(B\tau)}_{w_1(\tau)} \underbrace{\sqrt{T}\text{sinc}(T\nu)}_{w_2(\nu)}. 
\label{eq:sinc1}
\end{equation}
For sinc filter, the frame duration $T'=T$ and frame bandwidth $B'=B$ (i.e., there is no time or bandwidth expansion), resulting in a spectral efficiency of $\frac{BT}{B'T'}=1$ symbol/dimension. 

{\em RRC filter:}
For RRC filter, $w_\text{tx}(\tau,\nu)$ is given by
\begin{equation}
w_\text{tx}(\tau,\nu) = \underbrace{\sqrt{B} \ \text{rrc}_{\beta_\tau}(B\tau)}_{w_1(\tau)} \ \underbrace{\sqrt{T} \ \text{rrc}_{\beta_\nu}(T\nu)}_{w_2(\nu)},
\label{eq:rrc1}
\end{equation}
where $0\leq \beta_{\tau},\beta_\nu \leq 1$ and
\begin{equation}
\text{rrc}_{\beta}(x)=\frac{\sin \left(\pi x(1-\beta) \right)+4\beta x\cos \left(\pi x(1+\beta) \right)}{\pi x \left(1-(4\beta x)^2 \right)}. 
\end{equation}
It can be seen that the choice of $\beta_\tau=\beta_\nu=0$ in the RRC filter (\ref{eq:rrc1}) specializes to the sinc filter. Also, for $\beta_\nu>0$ and $\beta_\tau>0$, there is time and bandwidth expansion such that  $T'=T(1+\beta_{\nu})$ and $B'=B(1+\beta_{\tau})$, resulting in a spectral efficiency of $\frac{BT}{B'T'}<1$ symbol/dimension. 

{\em Gaussian filter:}
For Gaussian filter, $w_\text{tx}(\tau,\nu)$ is given by \cite{zak_otfs7}
\begin{equation}
\hspace{-2mm}
w_\text{tx}(\tau,\nu) \hspace{-0.5mm} = \hspace{-0.5mm} \underbrace{\left(\frac{2\alpha_{\tau}B^2}{\pi}\right)^{\frac{1}{4}}e^{-\alpha_{\tau}B^{2}\tau^{2}}}_{w_1(\tau)} \ \underbrace{\left(\frac{2\alpha_{\nu}T^2}{\pi}\right)^{\frac{1}{4}}e^{-\alpha_{\nu}T^{2}\nu^{2}}}_{w_2(\nu)}\hspace{-1mm}. \hspace{-0mm}
\label{eq:gauss1}
\end{equation}
The Gaussian pulse can be configured by adjusting the parameters $\alpha_{\tau}$ and $\alpha_{\nu}$. Because of the infinite support in Gaussian pulse, a time duration $T'$ where 99$\%$ of the frame energy is localized in the time domain and a bandwidth $B'$ where 99$\%$ of the frame 
energy is localized in the frequency domain are considered \cite{zak_otfs7}. No time and bandwidth expansion (i.e., $T'=T$, $B'=B$) in the Gaussian pulse corresponds to setting 
$\alpha_{\tau}=\alpha_{\nu}=1.584$. 

\vspace{-1mm}
\section{Delay-Doppler I/O Relation Estimation}
\label{sec3}
To perform the equalization/detection task at the receiver, knowledge of the DD I/O relation, i.e., the effective channel matrix ${\bf H_\text{eff}}$ in (\ref{sys_mod}), is needed. This can be obtained using two approaches, namely, model-dependent and model-free approaches \cite{zak_otfs2}.
In model-dependent approach, the parameters of the physical channel $h_\text{phy}(\tau,\nu)$, i.e., $\{\tau_i$, $\nu_i$, $h_i$\}s, are estimated using a channel estimation scheme and these estimated parameters are then used to construct the I/O relation. That is, use the estimated $\{\tau_i,\nu_i,h_i\}$s to compute $h_\text{eff}(\tau,\nu)$ defined in (\ref{cont1}) 
and sample it to obtain $h_\text{eff}[k,l]$ as in (\ref{discr2}), which when substituted in (\ref{eqn_channel_matrix}) gives the estimated I/O relation $\hat{\bf H}_\text{eff}$.

Model-free approach does not require explicit estimation of the physical channel parameters
$\{\tau_i,\nu_i,h_i\}$s. Instead, the I/O relation can be obtained by sending a pilot symbol in a frame and directly reading out the corresponding DD domain output samples in $\mathcal{D}_0$ at the receiver. Due to its simplicity, we consider model-free approach for I/O relation estimation, which is presented below for exclusive and embedded pilot frames. 

\vspace{-2mm}
\subsection{Exclusive pilot frame}
For a pilot at the origin (0,0), the channel response consists of two terms, the first term due to the pilot at the origin and the other term due to its quasi-periodic replicas, i.e., the response is given by 
\begin{equation}
h_\text{eff}[k,l] + \sum_{n,m\in\mathbb{Z},\newline (n,m)\neq(0,0)}\hspace{-6mm}h_\text{eff}[k-nM,l-mN]e^{j2\pi\frac{nl}{N}}.
\label{c_rsp}
\end{equation}
In the crystalline regime of operation, where the maximum delay and Doppler spreads of the effective channel (denoted by $\tau_\text{max}$ and $\nu_\text{max}$, respectively) are less than the delay and Doppler periods, respectively (i.e., $\tau_{\mathrm{max}}<\tau_{\mathrm{p}}$ and $\nu_{\mathrm{max}}<\nu_{\mathrm{p}}$), the local responses do not interfere with each other significantly \cite{zak_otfs2}. Therefore, the model-free I/O relation estimation approach considers only the $(0,0)$th local response.  

In an exclusive pilot frame, a pilot $x_{\text{p}}[k,l]$ located at $(k_{\text{p}}, l_{\text{p}})=(M/2, N/2)$ and zeros at other locations is sent to estimate the effective channel $h_{\mathrm{eff}}[k,l]$. The channel response for this exclusive pilot frame is given by 
\begin{align}
y_{\text{p}}[k,l] = & \ h_{\mathrm{eff}}[k,l]*_{\sigma\text{d}}x_{\text{p}}[k,l] \nonumber \\
=&\sum_{m,n\in\mathbb{Z}}h_{\mathrm{eff}}[k-(k_{\text{p}}+nM),l-(l_{\text{p}}+mN)] \nonumber \\
&e^{j2\pi\frac{nl_{\text{p}}}{N}}e^{j2\pi\frac{(l-l_{\text{p}}-mN)(k_{\text{p}}+nM)}{MN}}.
\label{c_rsp2}
\end{align}
In the crystalline regime, the total response in the fundamental period coincides with the $(0,0)$th local response ($m=n=0$), given by 
\begin{equation}
y_{\text{p}}[k,l]=h_{\mathrm{eff}}[k-M/2,l-N/2]e^{j\pi\frac{\left(l-\frac{N}{2}\right)}{N}},
\end{equation}
for $0\leq k<M$ and $0\leq l<N$. Consequently, the effective channel estimate is obtained as 
\begin{eqnarray}
\hat{h}_{\mathrm{eff}}[k,l] \hspace{-0.5mm} = \hspace{-0.5mm}
\begin{cases}
y_{\text{p}}\left[k+\frac{M}{2},l+\frac{N}{2}\right]e^{-j\pi\frac{l}{N}}, & \hspace{-2mm} -\frac{M}{2}\leq k<\frac{M}{2}, \\
& \hspace{-2mm} -\frac{N}{2}\leq l<\frac{N}{2}, \\ 
0, \ \ \mathrm{otherwise}.
\end{cases}
\hspace{-1mm}
\label{eqn:href_est}
\end{eqnarray}
The above estimated coefficients $\hat{h}_{\mathrm{eff}}[k,l]$ are used in (\ref{eqn_channel_matrix}) to obtain the $\hat{\bf H}_\text{eff}$. Note that the accuracy of this estimate in terms of normalized mean squared error (MSE), defined as the average of $\frac{||\bf{H}_{\text{eff}}-\hat{\bf H}_{\text{eff}}||_{F}^{2}}{||\bf{H}_{\text{eff}}||_{F}^{2}}$, is influenced by the choice of the filter, particularly the side lobe characteristics of the filter. Lower the side lobe levels, better will be the accuracy, because lower side lobes result in a weak second term in the channel response due to the replicas (see Eq. (\ref{c_rsp})). 

\vspace{0mm}
\begin{figure}
\hspace{2mm}
\includegraphics[width=9cm,height=6cm]{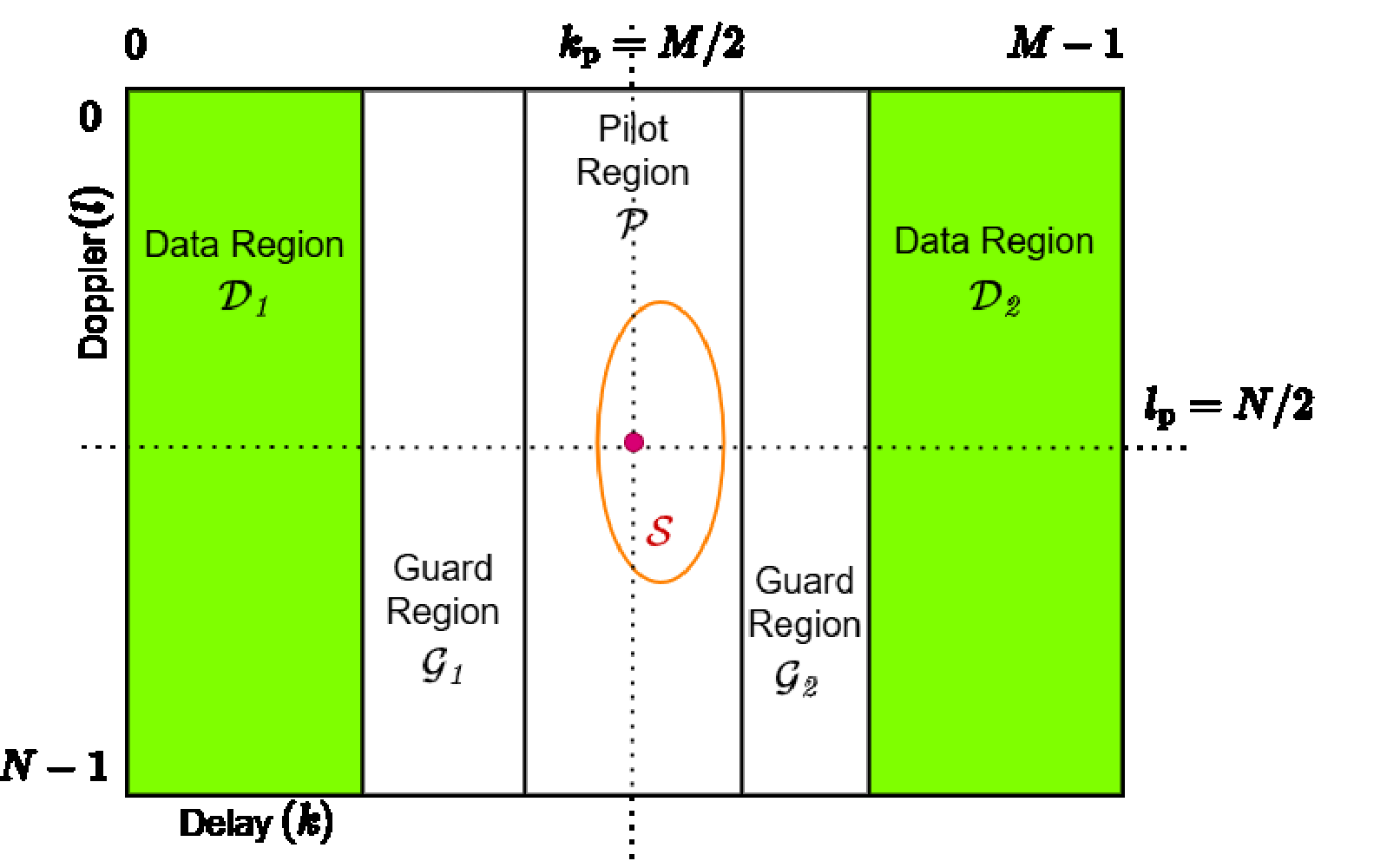}
\caption{Embedded pilot frame with pilot symbol, pilot region, guard region, and data region.}
\label{fig:embedded_pilot}
\vspace{-5mm}
\end{figure}

\vspace{-4mm}
\subsection{Embedded pilot frame}
We consider the embedded pilot frame shown in Fig. \ref{fig:embedded_pilot} \cite{zak_otfs7}. It consists of a pilot symbol located at $(k_{\text{p}}, l_{\text{p}})=(M/2,N/2)$, a data region $\mathcal{D}=\mathcal{D}_1\cup \mathcal{D}_2$ in which data symbols are transmitted, and a region in between (pilot region $\mathcal{P}$ + guard region $\mathcal{G}=\mathcal{G}_1\cup \mathcal{G}_2$) where no symbols are transmitted. 
The support of the effective channel, denoted by $\mathcal{S}$, is marked/represented by the ellipse in Fig. \ref{fig:embedded_pilot}. The pilot region is designed to encompass $\mathcal{S}$, and the guard regions act as buffers between the pilot and data regions to mitigate interference between them.
The pilot region spans from $k_{\text{p}}-p_1$ to $k_{\text{p}}+k_{\max}+p_2$, and the guard region is defined by the boundaries $k_{\text{p}}-k_{\max}-g_1$ and $k_{\text{p}}+k_{\max}+g_2$. Here, $k_{\max} = \lceil B\max(\tau_{i})  \rceil$ represents the maximum delay spread of the physical channel and $p_1, p_2, g_1, g_2$ are non-negative integers. The additional bins within these regions represented by $p_1, p_2, g_1, g_2$ accommodate the signal spread caused by the pulse shaping filters and can be chosen according to the system bandwidth.  

For $0\leq k\leq M-1$ and $0\leq l\leq N-1$, the symbol $x[k,l]$ in the frame is given by
\begin{eqnarray}
x[k,l]=
\begin{cases}
\sqrt{E_\text{p}}, \quad \quad \quad \quad \ (k,l)=(k_\text{p},l_\text{p}), \\
\sqrt{\frac{E_\text{d}}{|\mathcal{D}|}}x_{\text{d}}[k,l], \quad \ (k,l)\in \mathcal{D}, \\
0, \qquad \qquad \qquad \  \mathrm{otherwise,}
\end{cases}
\label{eqn:noise_integral_closed}
\end{eqnarray}
where $x_\text{d}[k,l]$ is the information symbol at location $(k,l)$. Taking $\mathbb{E}[|x_\text{d}[k,l]|^{2}]=1$, the average energy transmitted in a frame is $E_\text{p}+E_\text{d}$ and the average transmitted power is $(E_\text{p}+E_\text{d})/T'$. Normalizing the channel gains as $\sum_{i=1}^{P}\mathbb{E}[|h_i|^{2}]=1$, the data SNR is given by
$\gamma_{\text{d}}=\frac{E_\text{d}}{N_{0}B'T'}$ and the pilot SNR is given by
$\gamma_{\text{p}}=\frac{E_\text{p}}{N_{0}B'T'}$. The term  $E_{\text{p}}/E_{\text{d}}$ is the ratio of the pilot power to data power ratio (PDR).

In each frame, the  effective channel coefficients $\{h_{\mathrm{eff}}[k,l]\}$ are estimated based on the received pilot symbols at locations within the pilot region ${\mathcal{P}}$ using (\ref{eqn:href_est}). These estimates are used to construct the estimated effective channel matrix $\hat{\mathbf{H}}_{\text{eff}}$. Note that the estimation accuracy here is affected by the interference from data symbols (in addition to self-interference due to pilot symbol replicas and noise), which is determined by the pulse shape. The received DD symbols $y_{\mathrm{dd}}[k,l],\ (k,l)\in \mathcal{D} \hspace{0.5mm} \cup  \hspace{0.5mm} \mathcal{G}$ are arranged as a vector of length $MN-|\mathcal{P}|$, which is the vector of the $|\mathcal{D}|$ transmitted symbols times the effective channel matrix plus the noise vector \cite{zak_otfs7}. Information symbols are detected from the vector of received symbols in $\mathcal{D} \cup \mathcal{G}$. Note that, among other things, the equalizer/detection performance here is affected by the interference from pilot. 

\vspace{-1mm}
\section{Proposed Gaussian-Sinc DD filter} 
\label{sec4}
In the Zak-OTFS literature, sinc, RRC, and Gaussian pulse shaping filters have been considered. In this section, we present the rationale for a new pulse shaping filter for Zak-OTFS, the proposed Gaussian-sinc (GS) filter, and the derivation of closed-form expressions for the I/O relation and noise covariance with the proposed GS filter.  

\begin{figure}
\centering
\includegraphics[width=9cm,height=6cm]{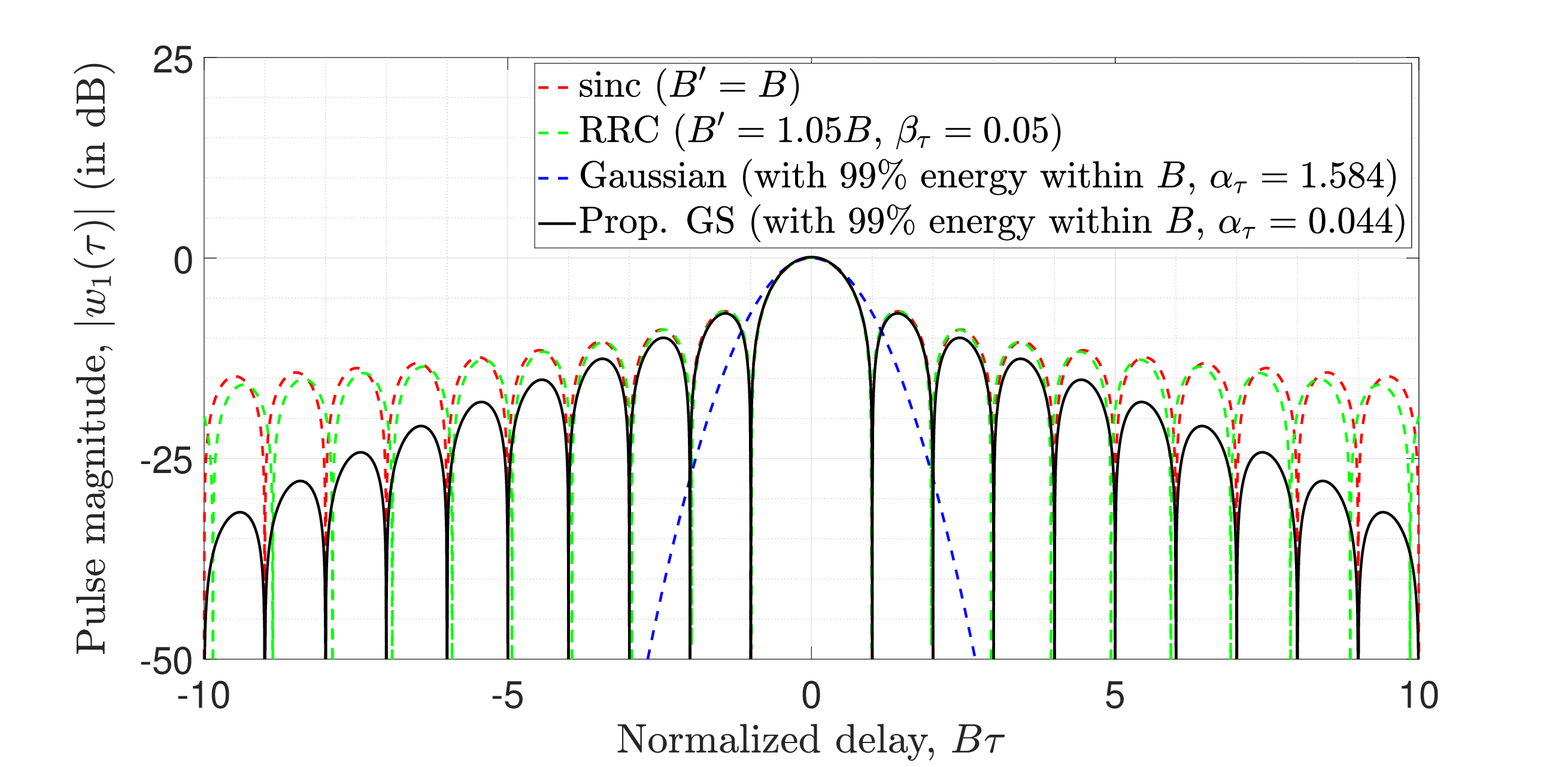}
\caption{Delay pulse magnitude $|w_{1}(\tau)|$ (in dB) as a function of the normalized delay $B\tau$.}
\label{pulse_shapes}
\end{figure}

\vspace{-3mm}
\subsection{Rationale for a new filter}
\begin{figure*}
\hspace{2mm}
\subfloat[BER vs SNR with perfect CSI]  {\includegraphics[width=6cm,height=4.5cm]{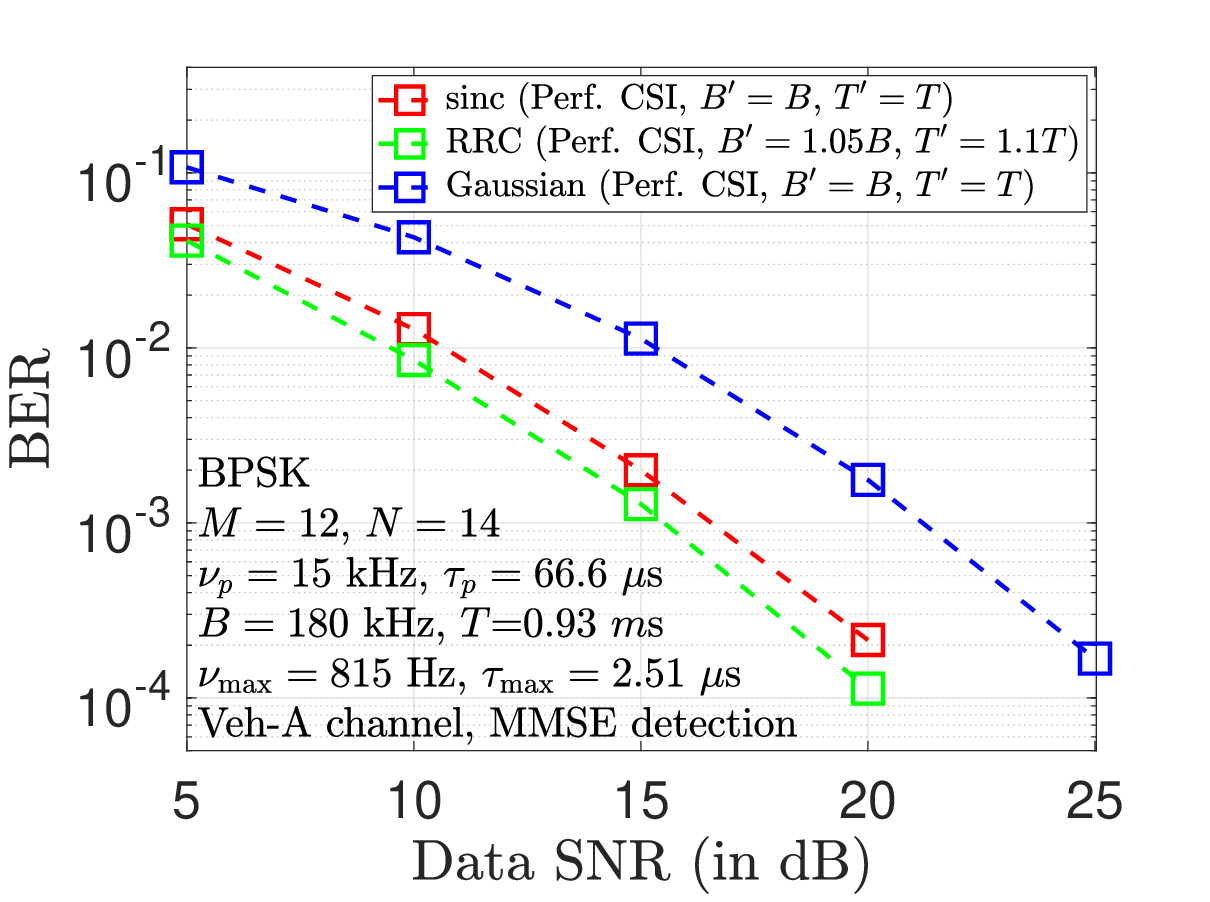} \label{fig:motiv_a}}\hfill  \subfloat[MSE vs SNR with embedded pilot]
{\includegraphics[width=6cm, height=4.5cm]{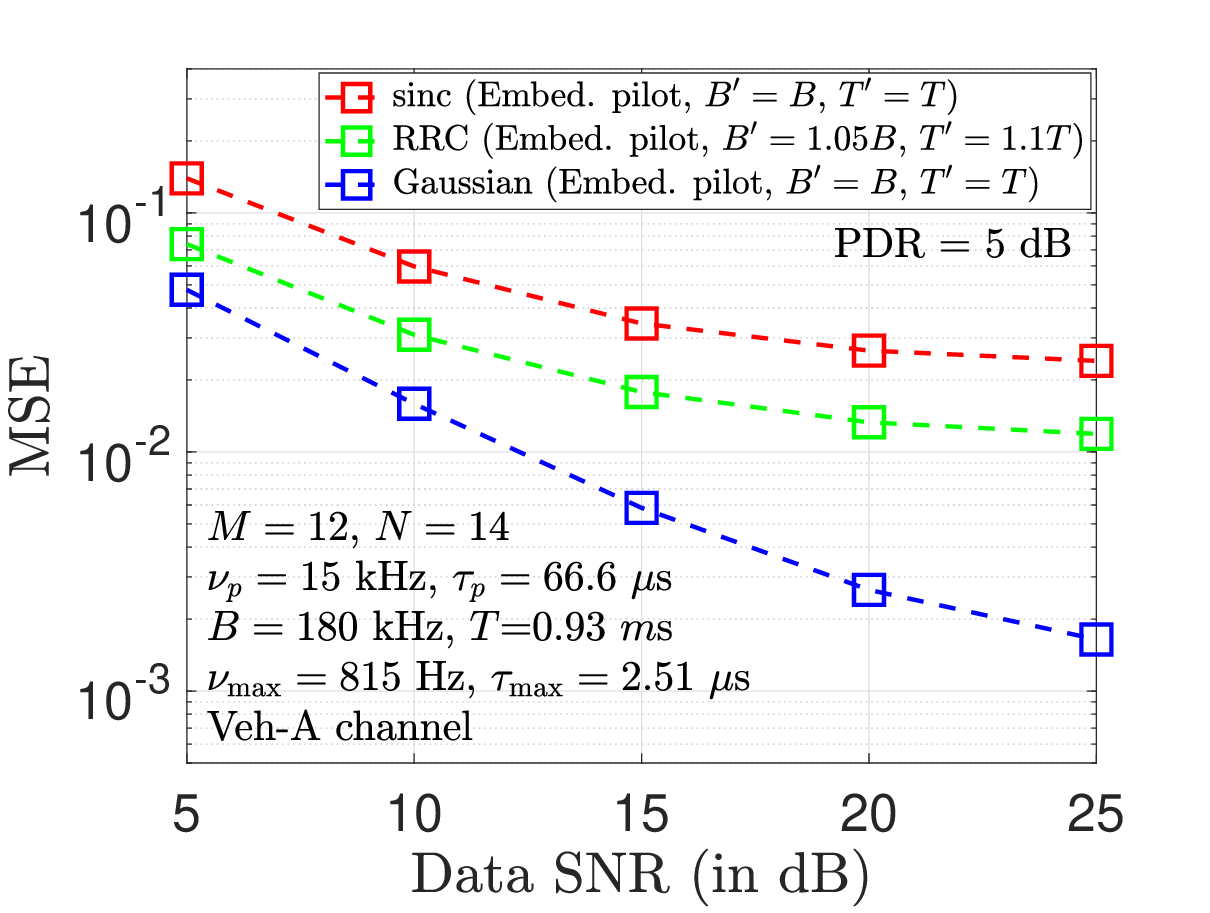} \label{fig:motiv_b}}
\subfloat[BER vs SNR with embedded pilot]
{\includegraphics[width=6cm, height=4.5cm]{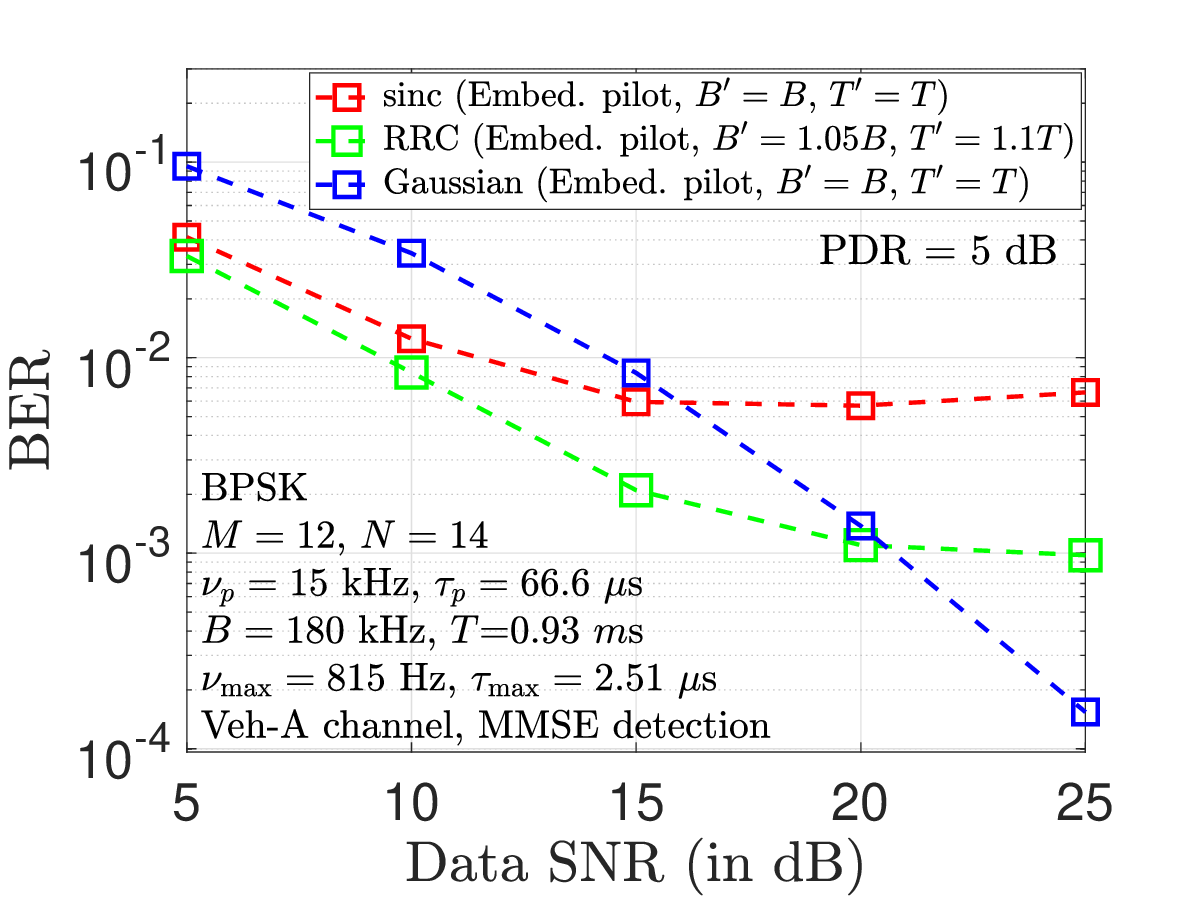} \label{fig:motiv_c}}
\caption{Performance of sinc and Gaussian filters (a) with perfect CSI and (b),(c) with model-free I/O relation estimation.}
\vspace{-4mm}
\label{fig:motiv}
\end{figure*}

In Fig. \ref{pulse_shapes}, we plot the delay pulse magnitude $|w_1(\tau)|$ in dB scale as a function of the normalized delay $B\tau$ for sinc, RRC, and Gaussian filters. Similar characteristics can be observed for Doppler pulse magnitude $|w_2(\nu)|$ as a function of the normalized Doppler $T\nu$. The sinc filter has ideal main lobe characteristics with nulls at the Nyquist sampling points on the DD grid (i.e., at $\tau=\frac{k}{B}$, $\nu=\frac{l}{T}$, $k,l \in \mathbb{Z}\backslash 0$). But it has the drawback of high side lobe levels. The RRC filter alleviates the issue of high side lobes in sinc filter through the choice of $\beta_\tau$ and $\beta_\nu$ parameters. But this is achieved at the expense of increased time and bandwidth, since $T'=T(1+\beta_\nu)$, $B'=B(1+\beta_\tau)$, and $0< \beta_\tau,\beta_\nu \leq 1$ . The Gaussian filter has very low side lobe levels, but it does not have good main lobe characteristics. In particular, it does not have nulls at the Nyquist sampling points. Instead, it has a high value closer to the peak value. These varied characteristics of the sinc, RRC, and Gaussian filters affect the receiver performance in different ways. For example, presence of nulls at the Nyquist points positively influences the equalization/detection performance, while having high non-zero values at these points has a negative influence on the equalization/detection performance. Likewise, very low side lobes positively influences the I/O relation estimation performance, while high side lobes influences it negatively. We illustrate the above points through the performance plots in Figs. \ref{fig:motiv_a}, \ref{fig:motiv_b}, \ref{fig:motiv_c}.
\begin{table}
\centering
\begin{tabular}{|c|c|c|c|c|c|c|}
\hline
Path index ($i$)         & 1 & 2    & 3    & 4    & 5    & 6    \\ \hline
Delay $\tau_{i}$ ($\mu s$)      & 0 & 0.31 & 0.71 & 1.09 & 1.73 & 2.51 \\ \hline
Relative power 
(dB) & 0 & -1   & -9   & -10  & -15  & -20  \\ \hline
\end{tabular}
\caption{Power delay profile of Veh-A channel model.}
\label{tab_pdp}
\vspace{-5mm}
\end{table}

For generating the performance plots in Figs. \ref{fig:motiv_a}, \ref{fig:motiv_b}, \ref{fig:motiv_c}, the following system parameters are used. A Zak-OTFS system with $M=12$, $N=14$, and BPSK is considered. The Doppler period taken to be $\nu_{\mathrm p}=15$ kHz. Therefore, the delay period is $\tau_{\mathrm p}=\frac{1}{\nu_{\mathrm p}}=66.66\ \mu$s. Consequently, the time duration of a Zak-OTFS frame is $T=N\tau_{\mathrm p}=0.93$ ms and the bandwidth is $B=M\nu_{\mathrm p}=180$ kHz. The receive filter is matched to the transmit filter \cite{zak_otfs6}-\cite{zak_otfs11}, i.e., 
\begin{equation}
w_{\mathrm{rx}}(\tau,\nu)=w^{*}_{\mathrm{tx}}(-\tau,-\nu) e^{j2\pi \nu\tau}.
\label{mat_filtering}
\end{equation}
The Veh-A fractional DD channel model \cite{ITU_VehA} having $P=6$ channel paths whose power delay profile is shown in Table \ref{tab_pdp} and a maximum Doppler shift of $\nu_{\max}=815$ Hz is considered. The Doppler shift of the $i$th path is modeled as $\nu_{i}=\nu_{\mathrm{max}}\cos\theta_{i},i=1,\ldots,P$, where $\theta_{i}$s are independent and uniformly distributed in $[0,2\pi)$. The considered $\tau_p,\nu_p$ values and channel spreads satisfy the crystallization condition. Minimum mean square error (MMSE) detection is used. For RRC filter, $\beta_\tau=0.05$ and $\beta_\nu=0.1$ are used. For Gaussian filter, $\alpha_\tau$ and $\alpha_\nu$ are taken to be 1.584. 

First, let us see how the choice of the filter affects the equalization/detection performance at the receiver. For this, we assume perfect channel state information (CSI). Figure \ref{fig:motiv_a} shows the BER performance of Zak-OTFS using sinc, RRC, and Gaussian filters with perfect CSI. We observe that the sinc filter achieves nearly 5 dB better performance compared to Gaussian filter. Note that, because of the perfect CSI assumption, there is no effect of I/O relation estimation on the detection performance. Consequently, the better performance of sinc filter with perfect CSI is attributed to the fact that it has nulls at Nyquist sampling points (leaving only a weak influence by the physical channel spread), whereas Gaussian filter has a high non-zero value at the $\tau=\frac{1}{B}$, $\nu=\frac{1}{T}$ sampling points (as per Fig. \ref{pulse_shapes}, this value is just 7 dB below the main lobe peak), which leads to high inter-symbol interference. With bandwidth and time expansion, RRC filter achieves slightly better performance compared to sinc filter performance.

Now, let us see how the filters affect performance when there is no perfect CSI assumption and a model-free I/O relation estimation scheme is used with embedded pilot frame. An embedded pilot frame structure shown in Fig. \ref{fig:embedded_pilot} with a PDR of 5 dB is considered.
As in Fig. \ref{fig:embedded_pilot}, the pilot is located at $(k_{\text{p}},l_{\text{p}})=(M/2,N/2)$  and the embedded frame parameters are fixed as $p_1=p_2=1$, $g_1=1$, $g_2=2$,  and $k_{\text{max}}=\lceil B\max(\tau_{i}) \rceil=1 $. Figure \ref{fig:motiv_b} shows the MSE performance of I/O relation estimation using sinc, RRC, and Gaussian filters. It is interesting to observe that while Gaussian filter's BER performance with perfect CSI is the worst (Fig. \ref{fig:motiv_a}), its MSE performance of I/O relation estimation is the best (Fig. \ref{fig:motiv_b}). This is attributed to the Gaussian filter's very low side lobes compared to those of sinc and RRC filters (see Fig. \ref{pulse_shapes}), which help to isolate the influence of interference from data/pilot replicas on estimation. However, for sinc filter, because of its high side lobes and consequent high interference levels, the MSE floors at a high value.  Figure \ref{fig:motiv_c} shows the BER performance comparison corresponding to the MSE performance comparison in Fig. \ref{fig:motiv_b}. From Fig. \ref{fig:motiv_c}, it is seen that, though Gaussian filter performs better than sinc filter in terms of MSE, there is a cross-over in their BER performance. This can be explained as follows. Because of its very good I/O relation estimation, the Gaussian filter's BER for perfect CSI and estimated CSI are very close (see BER plots of Gaussian filter in Figs. \ref{fig:motiv_a} and \ref{fig:motiv_c}). Whereas, because of its poor I/O relation estimation, the sinc filter's BER degrades significantly and floors at high SNRs, where the MSE floor (due to high data/pilot replicas' interference and pilot-data interference) dominates BER performance over noise variance. At low SNRs, the sinc filter has the advantage of low inter-symbol interference due to its nulls, whereas Gaussian filter suffers from high inter-symbol interference because of its poor main lobe characteristics leading to its poorer performance compared to sinc filter. Also, RRC filter performs slightly better than sinc filter, and this comes at the cost of time and bandwidth expansion.    
 
The above observations indicate that the Gaussian and sinc filters have complementary merits with respect to I/O relation estimation and detection tasks. Therefore, a filter which possesses the merits of both without bandwidth or time expansion is of interest, and this forms the essence of the GS filter proposed in the following subsection.

\vspace{-3mm}
\subsection{Proposed GS filter}
The proposed GS filter aims to simultaneously achieve the complementary strengths of Gaussian filter (good I/O relation estimation) and sinc filter (good equalization/detection) without bandwidth or time expansion. Towards this, the proposed filter is devised in a separable form $w_\text{tx}(\tau,\nu)=w_1(\tau)w_2(\nu)$, where $w_1(\tau)$ is a product function in $\tau$ variable of the form    
\begin{equation}
w_{1}(\tau)=\Omega_{\tau}\sqrt{B}\mathrm{sinc}(B\tau)e^{-\alpha_{\tau}B^{2}\tau^{2}},
\label{delay_domain}
\end{equation}   
and $w_2(\nu)$ is a product function in $\nu$ variable of the form
\begin{equation}
w_{2}(\nu)=\Omega_{\nu}\sqrt{T}\mathrm{sinc}(T\nu)e^{-\alpha_{\nu}T^{2}\nu^{2}},
\label{Doppler_domain}
\end{equation}
so that the overall proposed filter is given by
\begin{equation}
w_\text{tx}(\tau,\nu)=\Omega_\tau\Omega_\nu\sqrt{BT} \mathrm{sinc}(B\tau)\mathrm{sinc}(T\nu)e^{-\alpha_{\tau}B^{2}\tau^{2}} e^{-\alpha_{\nu}T^{2}\nu^{2}}.
\label{gsf}
\end{equation}
Note that $w_1(\tau)$ and $w_2(\nu)$ are constructed as product of sinc and Gaussian shaping functions with energy normalization parameters $\Omega_\tau$ and $\Omega_\nu$. 
The parameters $\alpha_{\tau}$ and $\alpha_{\nu}$ fix the bandwidth $B$ and the time duration $T$, respectively, and 
the parameters $\Omega_{\tau}$ and $\Omega_{\nu}$ are used to normalize the energy of the filter to unity, i.e.,  
$\int |w_1(\tau)|^{2}d\tau =\int |w_{2}(\nu)|^{2} d\nu=1$.
The  expressions for $\Omega_\tau$ and $\Omega_\nu$ in terms of $\alpha_{\tau}$ and
$\alpha_{\nu}$, respectively, for unit energy normalization are obtained in Appendix \ref{appxA}. 

The delay pulse characteristics of the proposed GS filter is plotted in Fig. \ref{pulse_shapes} (along with those of sinc, RRC, and Gaussian filters). It can be seen that the GS filter retains the nulls of the sinc filter while reducing the side lobe levels without bandwidth and time expansion. The values of $\alpha_\tau$ and $\alpha_\nu$ in the proposed filter in (\ref{gsf}) for which there is no bandwidth and time expansion ($B'=B, T'=T$) and 99\% energy is contained within bandwidth $B$ and time duration $T$ are $\alpha_{\tau}=\alpha_{\nu}=0.044$, and the corresponding values of $\Omega_{\tau}$ and $\Omega_{\nu}$ are $\Omega_{\tau}=\Omega_{\nu}=1.0278$. 

\vspace{-2mm}
\subsection{Closed-form expressions for I/O relation/noise covariance} 
To facilitate performance analysis/simulation of Zak-OTFS with the proposed GS filter, here we derive closed-form expressions for the DD domain I/O relation and noise covariance with the proposed filter. A receive filter matched to the proposed filter (as per Eq. (\ref{mat_filtering})) is considered. The effective channel in the continuous DD domain 
can be written as 
\begin{eqnarray}
h_{\mathrm{eff}}(\tau,\nu) & \hspace{-2mm} = & \hspace{-2mm} w_{\mathrm{rx}}(\tau,\nu)*_{\sigma}h_{\mathrm{phy}}(\tau,\nu)*_{\sigma}w_{\mathrm{tx}}(\tau,\nu) \nonumber \\ 
& \hspace{-30mm} = & \hspace{-17mm} w_{\mathrm{rx}}(\tau,\nu)*_{\sigma}\left(\sum_{i=1}^{P}h_{i}\delta(\tau-\tau_{i})\delta(\nu-\nu_{i})\right)*_{\sigma}w_{1}(\tau)w_{2}(\nu) \nonumber    \\ 
& \hspace{-30mm} = & \hspace{-17mm} w_{1}^{*}(-\tau)w_{2}^{*}(-\nu)e^{j2\pi\nu\tau}*_{\sigma}\bigg(\sum_{i=1}^{P}h_{i}w_{1}(\tau-\tau_{i})w_{2}(\nu-\nu_{i}) \nonumber \\ 
& \hspace{-30mm} & \hspace{-17mm} e^{j2\pi\nu_{i}(\tau-\tau_{i})}\bigg) \nonumber \\
& \hspace{-30mm} = & \hspace{-17mm} \sum_{i=1}^{P}
\underbrace{\left(\int w_{1}^{*}(-\tau')w_{1}(\tau-\tau_{i}-\tau') e^{-j2\pi\nu_{i}\tau'}d\tau'\right)}_{\overset{\Delta}{=}I_{i}^{(1)}(\tau)} \nonumber \\
& \hspace{-35mm} & \hspace{-20mm} 
\underbrace{\left(\int \hspace{-1mm} w_{2}^{*}(-\nu')w_{2}(\nu-\nu_{i}-\nu')e^{j2\pi\nu'\tau}d\nu'\hspace{-0.5mm} \right)}_{\overset{\Delta}{=}I_{i}^{(2)}(\tau,\nu)}
\hspace{-0.5mm} h_{i}e^{j2\pi\nu_{i}(\tau-\tau_{i})}.
\label{eqn:channel_matched}
\end{eqnarray}
\vspace{0mm}
We note that a general expression for $h_{\mathrm{eff}}(\tau,\nu)$ for an arbitrary pulse-shaping filter in the matched filter configuration is presented in \cite{zak_otfs5} (see Eq. (57) in \cite{zak_otfs5}). This $h_{\mathrm{eff}}(\tau,\nu)$ expression in \cite{zak_otfs5} is given in a form of two separable integrals, corresponding to the auto-ambiguity functions of the time and frequency domain representations of the Doppler and delay domain components of the pulse-shaping filter, respectively. Observe that the integrals in (\ref{eqn:channel_matched}) are similar to those of the auto-ambiguity integrals in \cite{zak_otfs5}. Here, we further simplify the integrals in (\ref{eqn:channel_matched}) to closed-form for the proposed GS filter. Accordingly, we specialize $w_1(\tau)$ and $w_2(\nu)$ in (\ref{eqn:channel_matched}) with those of the GS filter given in (\ref{delay_domain}) and (\ref{Doppler_domain}), respectively, and obtain closed-form expression for $h_{\mathrm{eff}}(\tau,\nu)$ for the proposed GS filter (see Theorem \ref{Thm1} below and Appendix \ref{appxB}).
\begin{theorem} 
The DD domain effective channel $h_{\mathrm{eff}}(\tau, \nu)$ in closed-form for GS filter is given by
\begin{eqnarray}
\hspace{-6mm}
h_{\mathrm{eff}}(\tau,\nu) & \hspace{-2mm} = & \hspace{-2mm} \sum_{i=1}^{P}h_{i}e^{j2\pi\nu_{i}(\tau-\tau_i)} \nonumber \\
& \hspace{-2mm} & \hspace{-2mm}
\cdot\left(I_{i,1}^{(1)}(\tau)\mathbbm{1}_{\{\tau\neq\tau_i\}}+I_{i,2}^{(1)}(\tau)\mathbbm{1}_{\{\tau=\tau_i\}}\right) \nonumber \\
& \hspace{-2mm} & \hspace{-2mm} \cdot \left(I_{i,1}^{(2)}(\tau,\nu)\mathbbm{1}_{\{\nu\neq\nu_i\}}+I_{i,1}^{(2)}(\tau,\nu)\mathbbm{1}_{\{\nu=\nu_i\}}\right),
\label{eqn:GS_match_channel}
\end{eqnarray}
where 
$\mathbbm{1}_{\{.\}}$ denotes the indicator function, and  
$I_{i,1}^{(1)}(\tau)$, $I_{i,2}^{(1)}(\tau)$, $I_{i,1}^{(2)}(\tau,\nu)$, and $I_{i,2}^{(2)}(\tau,\nu)$ are defined in Appendix \ref{appxB}. 
\label{Thm1}
\end{theorem}
\vspace{0mm}
\vspace{-1mm}
\begin{IEEEproof}
See Appendix \ref{appxB}.
\end{IEEEproof}

Now, the continuous DD domain noise at the output is given by
\begin{eqnarray}
n_{\mathrm{dd}}^{w_{\mathrm{rx}}}(\tau,\nu) & \hspace{-2mm} = & \hspace{-2mm} w_{\mathrm{rx}}(\tau,\nu)*_{\sigma}n_{\mathrm{dd}}(\tau,\nu) \nonumber \\ 
& \hspace{-33mm} = & \hspace{-18mm} w_{1}^{*}(-\tau)w_{2}^{*}(-\nu)e^{j2\pi\nu\tau}\hspace{-1mm} *_{\sigma}\hspace{-0.5mm} \Big(\hspace{-0.5mm} \sqrt{\tau_{p}}\sum_{q\in\mathbb{Z}}n(\tau+q\tau_{p})e^{-j2\pi\nu q\tau_{p}}\Big) \nonumber \\ 
& \hspace{-33mm} = & \hspace{-18mm} \sqrt{\tau_{p}}\sum_{q=-\infty}^{\infty}e^{-j2\pi\nu q\tau_{p}}\left(\int w_{1}^{*}(-\tau')n(\tau-\tau'+q\tau_{p})d\tau'\right) \nonumber \\ 
& \hspace{-33mm} & \hspace{-18mm}  \underbrace{\left(\int w_{2}^{*}(-\nu')e^{j2\pi\nu'(\tau+q\tau_{p})}\right)}_{\overset{\Delta}{=}I_{q}^{(3)}(\tau)}.
\label{eqn:noise_matched}
\end{eqnarray}
A general expression for the noise covariance for an arbitrary pulse shape with the matched filter configuration is presented in \cite{zak_otfs5} (see the expression after Eq. (47) in page 4469 of \cite{zak_otfs5}). This noise covariance expression in \cite{zak_otfs5} consists of integrals in the form of auto-ambiguity function of the delay domain component 
of the pulse-shaping filter, and auto-correlation function of the Doppler domain component 
of the pulse-shaping filter. Here, using the DD domain noise in (\ref{eqn:noise_matched}), we derive the covariance of the noise in closed-form for the proposed GS filter (see Theorem \ref{Thm2} below and Appendix \ref{appxC}).

\begin{theorem}
For all $k_1,k_2=0,1,...,M-1, \ l_1,l_2=0,1,...,N-1$, the $(k_{1}N+l_1+1,k_{2}N+l_{2}+1)$th element of the noise covariance matrix is given by
\begin{eqnarray}
\hspace{-1mm}
\mathbbm{E}[n_{\mathrm{dd}}[k_1,l_1],n_{\mathrm{dd}}^{*}[k_2,l_2]] & \hspace{-2mm} = & \hspace{-2mm} \tau_p\sum_{q_1=-\infty}^{\infty}\sum_{q_2=-\infty}^{\infty}e^{j2\pi\frac{q_2l_2-q_1l_1}{N}} \nonumber \\
& \hspace{-53mm} & \hspace{-43mm} \cdot g\left(\frac{k_1\tau_p}{M}+q_1\tau_p\right)g^{*}\left(\frac{k_2\tau_p} {M}+q_2\tau_p\right) \nonumber \\
& \hspace{-53mm} & \hspace{-43mm} 
\cdot \Big(S_{\{k_1,k_2,q_1,q_2\}}^{(1)}\mathbbm{1}_{\{x_{\{k_1,k_2,q_1,q_2\}}\neq 0\}} 
+S^{(2)}\mathbbm{1}_{\{x_{\{k_1,k_2,q_1,q_2\}}= 0\}} \Big), \nonumber \\
\label{eqn:noise_expectation}
\end{eqnarray}
where $g(.)$, $x_{\{k_1,k_2,q_1,q_2\}}$, 
$S^{(1)}_{\{k_1,k_2,q_1,q_2\}}$, and $S^{(2)}$ are defined in Appendix \ref{appxC}.
\label{Thm2}
\end{theorem}
\vspace{-1mm}
\begin{IEEEproof}
See Appendix \ref{appxC}.
\end{IEEEproof}

\section{Results and Discussions}
\label{sec5}
In this section, we present the numerical results on the MSE and BER performance of Zak-OTFS for sinc, RRC, Gaussian, and GS filters with model-free I/O relation estimation using exclusive and embedded pilots. We consider a system with $M=32$, $N=48$, and fix the pilot location at $(k_{\mathrm{p}},l_{\mathrm{p}})=(M/2,N/2)$. The Doppler period is fixed at $\nu_{\mathrm p}=15$ kHz and the delay period is $\tau_{\mathrm p}=\frac{1}{\nu_{\mathrm p}}=66.66\ \mu$s. The time duration of a frame is $T=N\tau_{\mathrm p}=3.2$ ms and the bandwidth is $B=M\nu_{\mathrm p}=480$ kHz. 
Receive filter $w_\text{rx}(\tau,\nu)$ is matched to the transmit filter $w_\text{tx}(\tau,\nu)$ (see Eq. (\ref{mat_filtering})).
We consider the Veh-A channel model \cite{ITU_VehA} having $P=6$ paths with fractional DDs and a PDP as detailed in Table \ref{tab_pdp}. The maximum Doppler shift is $\nu_{\mathrm{max}}=815$ Hz, and the Doppler shift of the $i$th path is modeled as $\nu_{i}=\nu_{\mathrm{max}}\cos\theta_{i},i=1,\ldots,P$, where $\theta_{i}$s are independent and uniformly distributed in $[0,2\pi)$. Also, in the simulations, the range of values of $m$ and $n$ in (\ref{eqn_channel_matrix}) is limited to -1 to 1, and this is found to ensure an adequate support set of $h_{\mathrm{eff}}[k,l]$ that captures the channel spread accurately. BPSK and 8-QAM modulation alphabets are considered.
MMSE detection is used at the receiver. No bandwidth/time expansion ($B'=B, T'=T$) is considered for the filters except RRC filter. For RRC filter, an expanded bandwidth of $B'=1.05B$ $(\beta_{\tau}=0.05)$ and an expanded time duration of $T'=1.1T$ $(\beta_{\nu}=0.1)$ are considered.

\begin{figure}[!t]
\centering
\includegraphics[width=9.0cm,height=6.5cm]{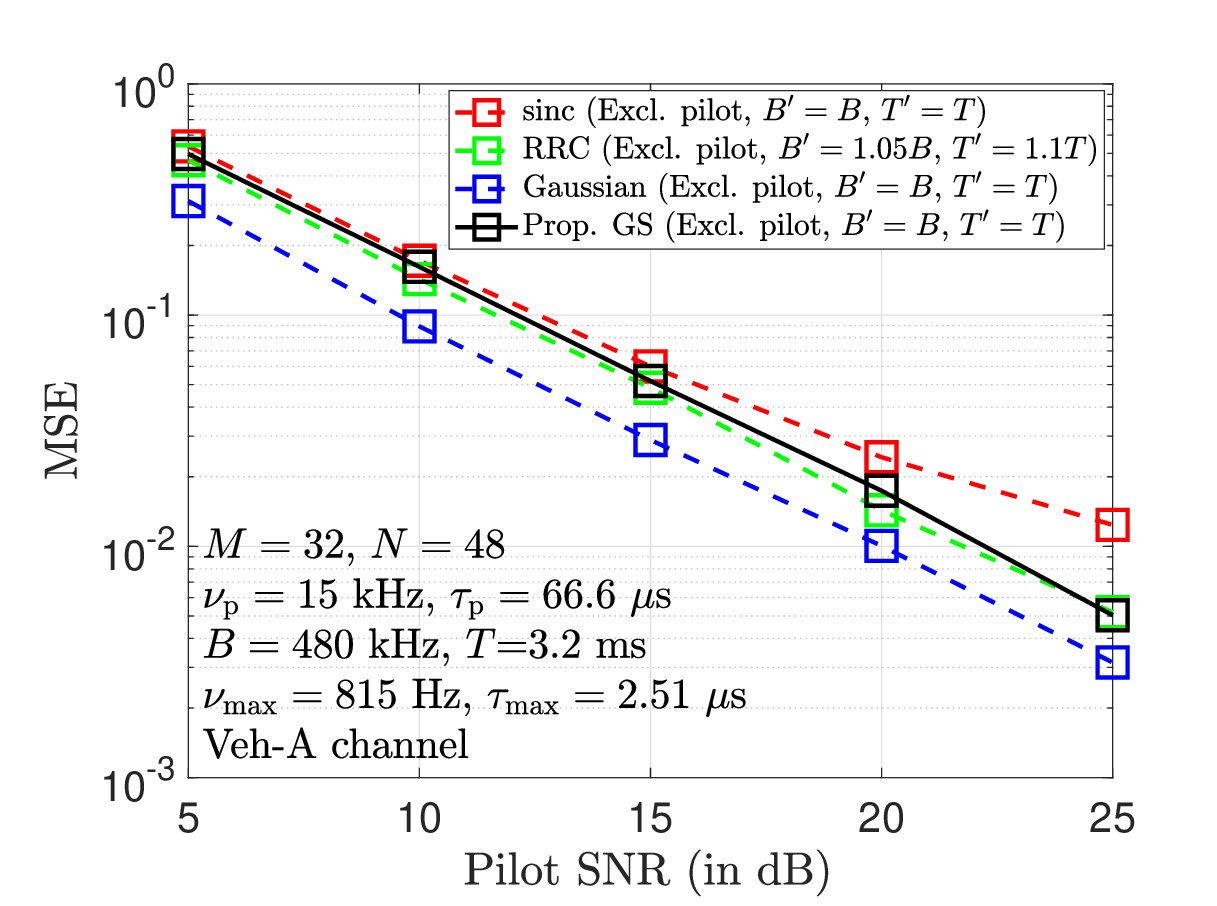}
\caption{MSE vs pilot SNR performance for different filters with exclusive pilot frame.}
\label{fig:mse_32X48_exclusive}
\vspace{-4mm}
\end{figure}

\vspace{-2mm}
\subsection{Performance with exclusive pilot frame}
\begin{figure}[!t]
\centering
\includegraphics[width=9.0cm,height=6.5cm]{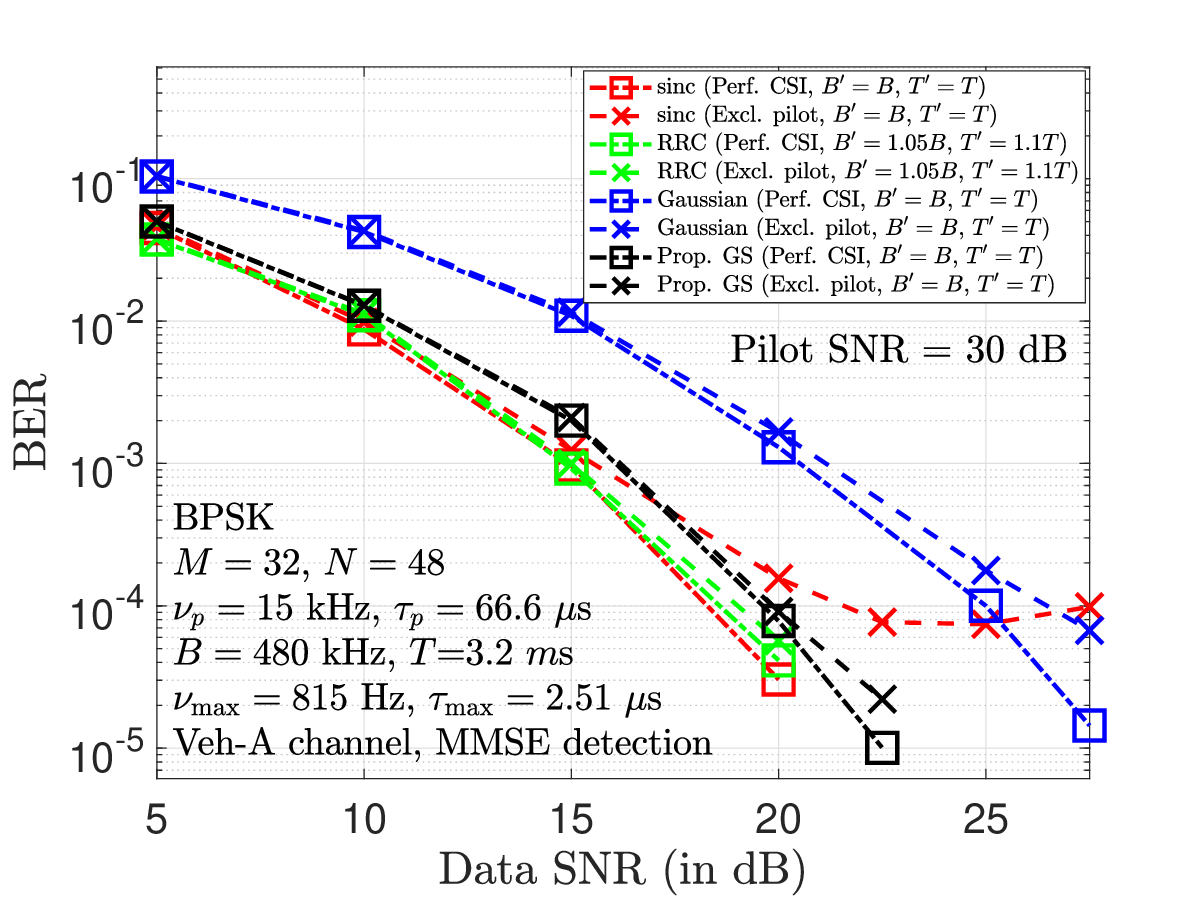}
\caption{BER vs data SNR performance of different filters with exclusive pilot frame at 30 dB pilot SNR. Perfect CSI performance is also shown.}
\label{fig:BER_32X48_exclusive}
\vspace{-4mm}
\end{figure}

Figures \ref{fig:mse_32X48_exclusive} 
and \ref{fig:BER_32X48_exclusive} show the MSE and BER performance of sinc, RRC, Gaussian, and the proposed GS filters using exclusive pilot frame. Figure 7 presents the effect of limited read-off from the exclusive pilot frame for I/O relation estimation.

\vspace{1mm}
\subsubsection{MSE performance}
Figure \ref{fig:mse_32X48_exclusive} shows the MSE performance as a function of pilot SNR. It is observed that the Gaussian filter performs better than the other three filters in terms of MSE performance. This characteristic is attributed to the highly localized nature  of the Gaussian filter with very low side lobes, resulting in negligible spread of the effective channel \big($h_{\mathrm{eff}}[k,l]$\big) outside the fundamental region $\mathcal{D}_{0}$, and this results in a very good estimate of the effective channel matrix $\hat{\textbf{H}}_{\mathrm{eff}}$. The MSE performance of the sinc filter is the poorest among all, which is due to its high side lobe levels that result in high effective channel spreads outside $\mathcal{D}_{0}$, leading to poor estimates. RRC filter performs slightly better than sinc filter, which is an artifact of the comparatively lower side lobe levels due to bandwidth and time expansion (refer Fig. \ref{pulse_shapes}). The proposed GS filter has low side lobe levels and achieves almost the same MSE performance as that of the RRC filter, but it achieves it without bandwidth and time expansion. 

\begin{figure}[!t]
\centering
\includegraphics[width=9.0cm,height=6.5cm]{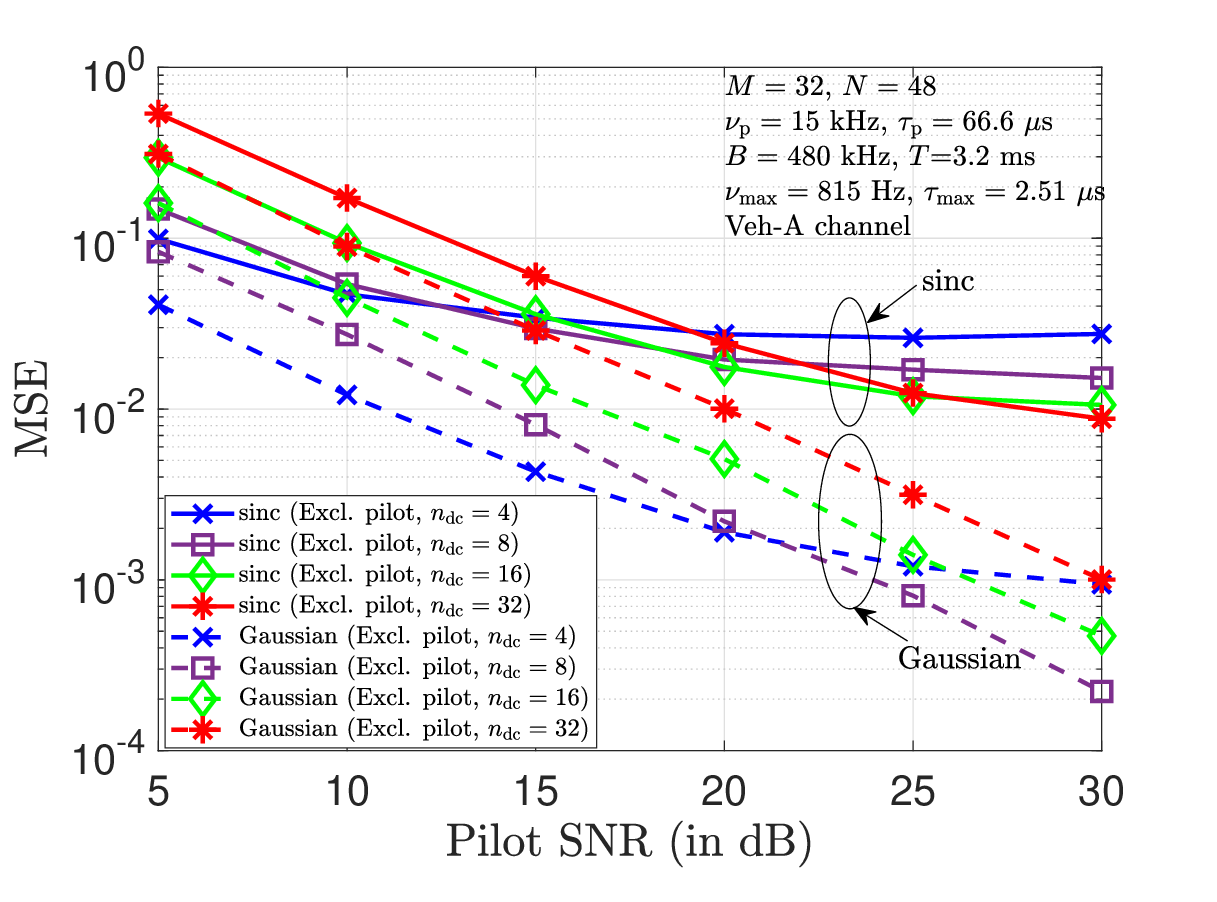}
\caption{MSE vs pilot SNR performance for sinc and Gaussian filters with limited read-off  in exclusive pilot frame.}
\label{fig:MSE_vs_SNR_different_delay_bins}
\vspace{-4mm}
\end{figure}

\vspace{1mm}
\subsubsection{BER performance}
Figure \ref{fig:BER_32X48_exclusive} shows the corresponding BER performance as a function of data SNR with BPSK at a pilot SNR of 30 dB. 
Performance with perfect CSI is also plotted for comparison. The sinc filter performs the best with perfect CSI, because its nulls at the sampling points cause weak inter-symbol interference among the data symbols which aids good detection performance. However, with I/O relation estimation using exclusive pilot, the presence of high side lobe levels and consequent high interference from pilot replicas results in a higher MSE (as seen in Fig. \ref{fig:BER_32X48_exclusive}), and this makes the BER to floor. RRC filter performs 
better due to time and bandwidth expansion. The Gaussian filter performs the worst with perfect CSI because of the absence of nulls at the sampling points, thereby causing high inter-symbol interference which results in poor data detection performance. However, its highly localized pulse shape results in a very accurate I/O relation estimation, and hence its performance with estimated CSI closely follows its own perfect CSI performance.
With I/O relation estimation, the proposed GS filter strikes a good balance between estimation and detection performance (with its low side lobes and nulls at sampling points) and achieves very good BER performance. 

\begin{figure}[!t]
\centering
\includegraphics[width=9.0cm,height=6.5cm]{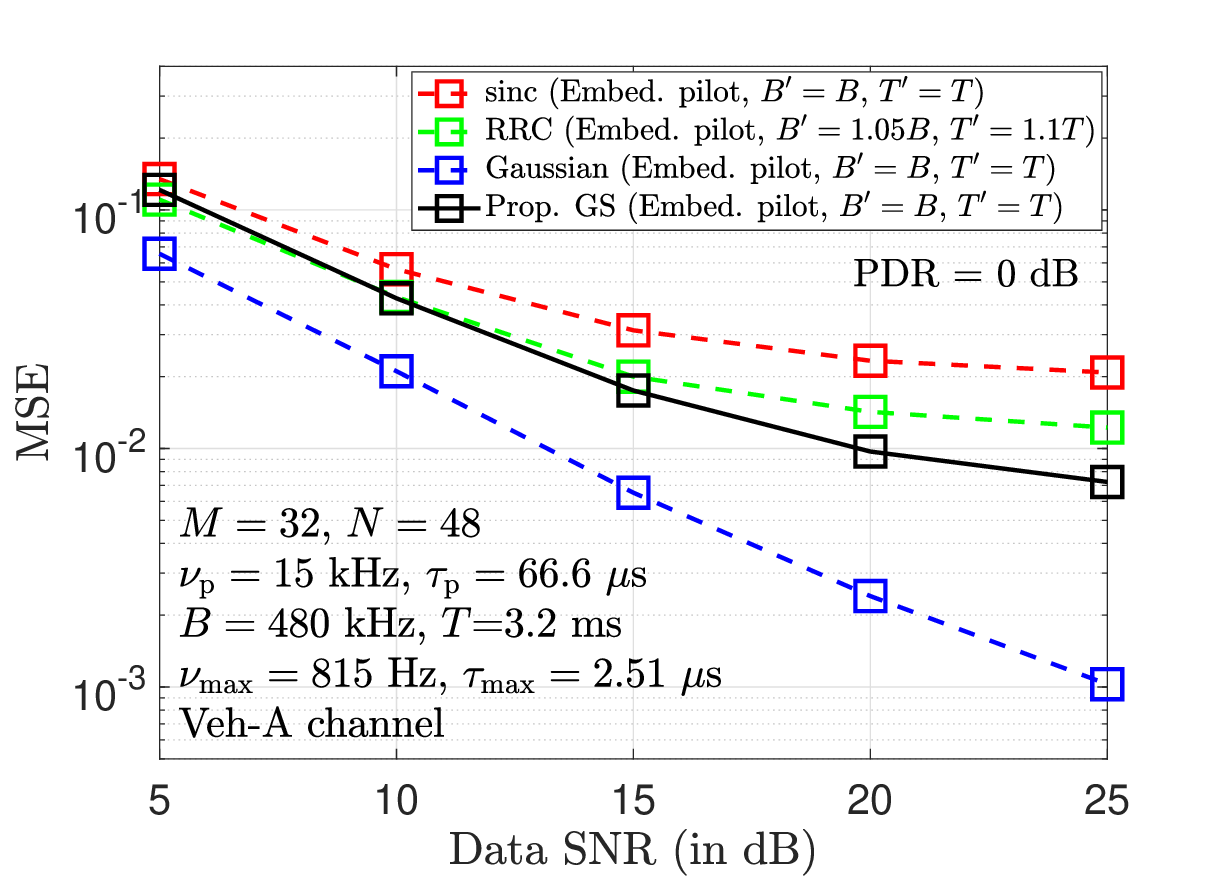}
\caption{MSE vs data SNR performance of different filters with embedded pilot frame at 0 dB PDR.}
\label{fig_mse}
\vspace{-4mm}
\end{figure}

\vspace{3mm}
\subsubsection{Effect of limited read-off for I/O relation estimation}
In model-free I/O relation estimation, we estimate the effective channel coefficients by reading off the received samples (see Eq. (\ref{eqn:href_est})) which are used in the summation for constructing the effective channel matrix (see Eq. (\ref{eqn_channel_matrix})). Let $n_{\mathrm{dc}}$ denote the number of delay columns around the pilot considered in the read-off. This determines the support of the estimated effective channel.  Note that, $n_{\mathrm{dc}}=M$ is used in exclusive pilot-based estimation, i.e., the entire received frame is read off.  
In embedded pilot-based estimation, the received samples are read off only from the pilot region, i.e.,  $n_{\text{dc}}<M$. Figure \ref{fig:MSE_vs_SNR_different_delay_bins} presents an assessment of the effect of  $n_{\mathrm{dc}}<M$ (i.e., limited read-off) in exclusive pilot-based estimation. The figure shows the MSE performance for sinc and Gaussian filters for $n_{\mathrm{dc}}=M, M/2,M/4,M/8$. 
We observe that $n_{\text{dc}}$ affects the MSE performance differently at low and high SNR regions. That is, at low SNRs, 
smaller $n_{\mathrm{dc}}$ gives better MSE, whereas, at high SNRs, 
larger $n_\text{dc}$ provides better MSE. This is because a smaller $n_\text{dc}$ means a fewer terms to be summed up to obtain the estimate of $\textbf{H}_{\mathrm{eff}}$ as per (\ref{eqn_channel_matrix}), 
which reduces the effect of noise on this estimate at low SNRs (where noise is dominant), leading to better MSE. Whereas, a larger $n_\text{dc}$ incorporates more number of terms in (\ref{eqn_channel_matrix}) in the construction of $\textbf{H}_{\mathrm{eff}}$, which, at high SNRs (where signal terms are dominant), is beneficial to achieve better MSE.

\vspace{-2mm}
\subsection{Performance with embedded pilot frame}
Here, we present the MSE and BER performance with embedded pilot frame for different filters. As shown in Fig. \ref{fig:embedded_pilot}, the pilot symbol is placed at $(k_{\text{p}},l_{\text{p}})=(M/2,N/2)$ and the embedded pilot frame parameters are fixed as $p_1=3$, $p_2=1$, $g_1=2$, $g_2=3$,  and $k_{\text{max}}=\lceil B\max(\tau_{i}) \rceil=2$.

\vspace{1mm}
\subsubsection{MSE performance}
In Fig. \ref{fig_mse}, the MSE performance of I/O relation estimation is plotted as a function of data SNR at a fixed PDR of 0 dB. It can be observed that the Gaussian filter consistently demonstrates the highest estimation accuracy, followed by the proposed GS filter, the RRC filter, and finally the sinc filter. Two sources of interference arise due to the higher side lobe levels inherent in non-Gaussian filters: 1) aliasing due to interference from quasi-periodic replicas which is self-interaction and 2) pilot-data interference within the same frame. While aliasing affects both exclusive and embedded pilot scenarios, pilot-data interference is an additional challenge specific to the embedded pilot setting. This combined interference significantly degrades the estimation accuracy of non-Gaussian filters, particularly at higher SNRs where the impact of noise diminishes and these interference effects become more prominent. This manifests as flooring in the MSE performance at high SNRs, and a wider performance gap emerges between Gaussian and non-Gaussian filters. Comparing the MSE performance of exclusive and embedded pilots in Figs. \ref{fig:mse_32X48_exclusive} and \ref{fig_mse}, respectively, we see a similar trend of MSE performance at low and high SNRs reported in Fig. \ref{fig:MSE_vs_SNR_different_delay_bins}, i.e., at low SNRs, embedded pilot-based estimation is better because of the small $n_{\mathrm{dc}}$ for the read-off.

\begin{figure}[!t]
\centering
\includegraphics[width=9.0cm,height=6.5cm]{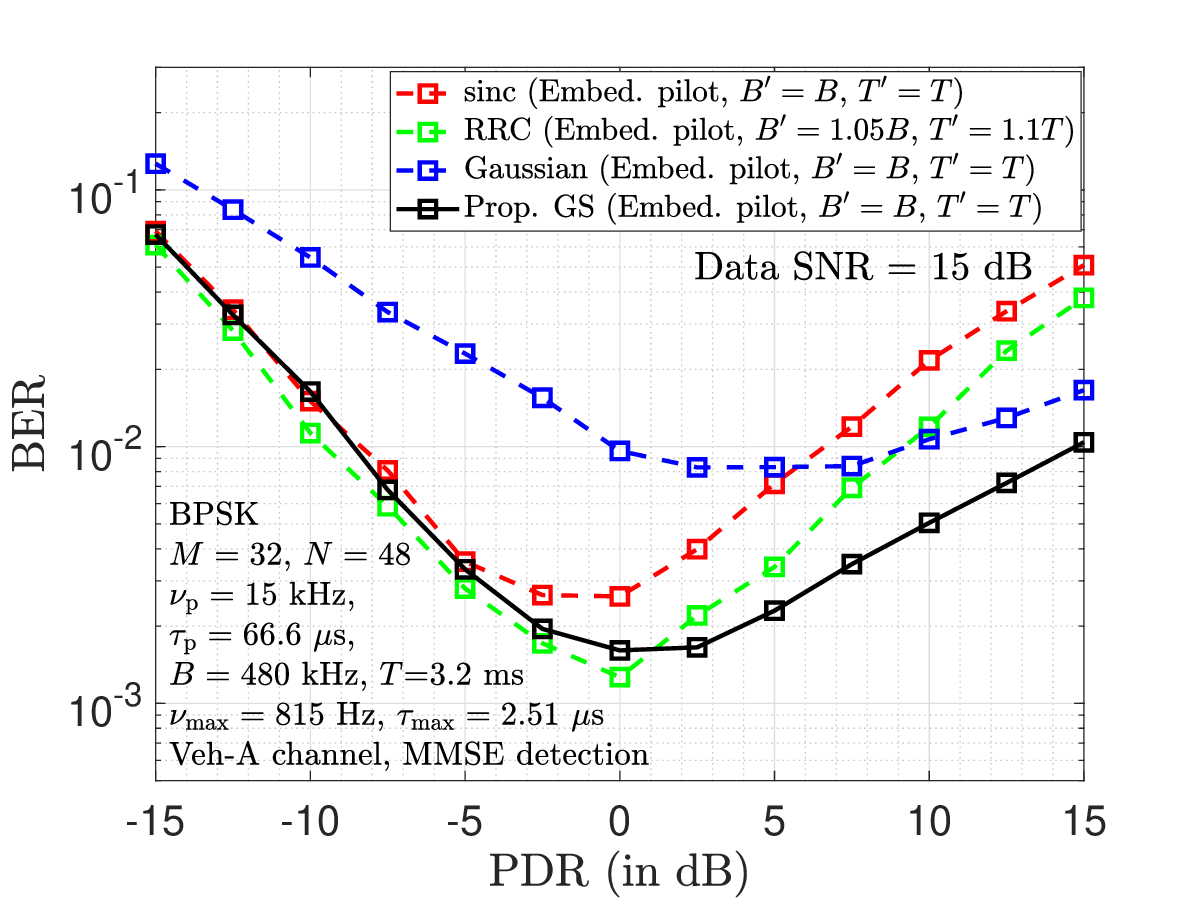}
\caption{BER vs PDR performance of different filters with embedded pilot frame at 15 dB data SNR.}
\label{fig_ber_pdr}
\vspace{-4mm}
\end{figure}

\vspace{1mm}
\subsubsection{BER performance}
Figure \ref{fig_ber_pdr} shows the BER performance of different filters as a function of PDR at a fixed data SNR of 15 dB. 
The BER curves exhibit U-shaped characteristics with respect to PDR. At low PDRs, poor estimation due to low pilot SNR degrades data detection and increases BER.  The BER improves with increase in PDR due to more accurate estimation. Conversely, excessive pilot power at high PDRs leads to significant pilot-data interference more than the noise effect, particularly more detrimental to non-Gaussian filters due to their higher side lobe levels. This manifests in a steeper increase in BER for non-Gaussian filters compared to Gaussian filter at high PDR values, indicating a higher sensitivity to strong pilot signals. The Gaussian filter, with its better DD localization and lower side lobe levels, is less susceptible to this interference, resulting in a more gradual increase in BER at high PDRs. The proposed GS filter consistently demonstrates the best performance because of its pulse shape which balances the complementary strengths of Gaussian and sinc filters.

\begin{figure}[!t]
\centering
\includegraphics[width=9.0cm,height=6.5cm]{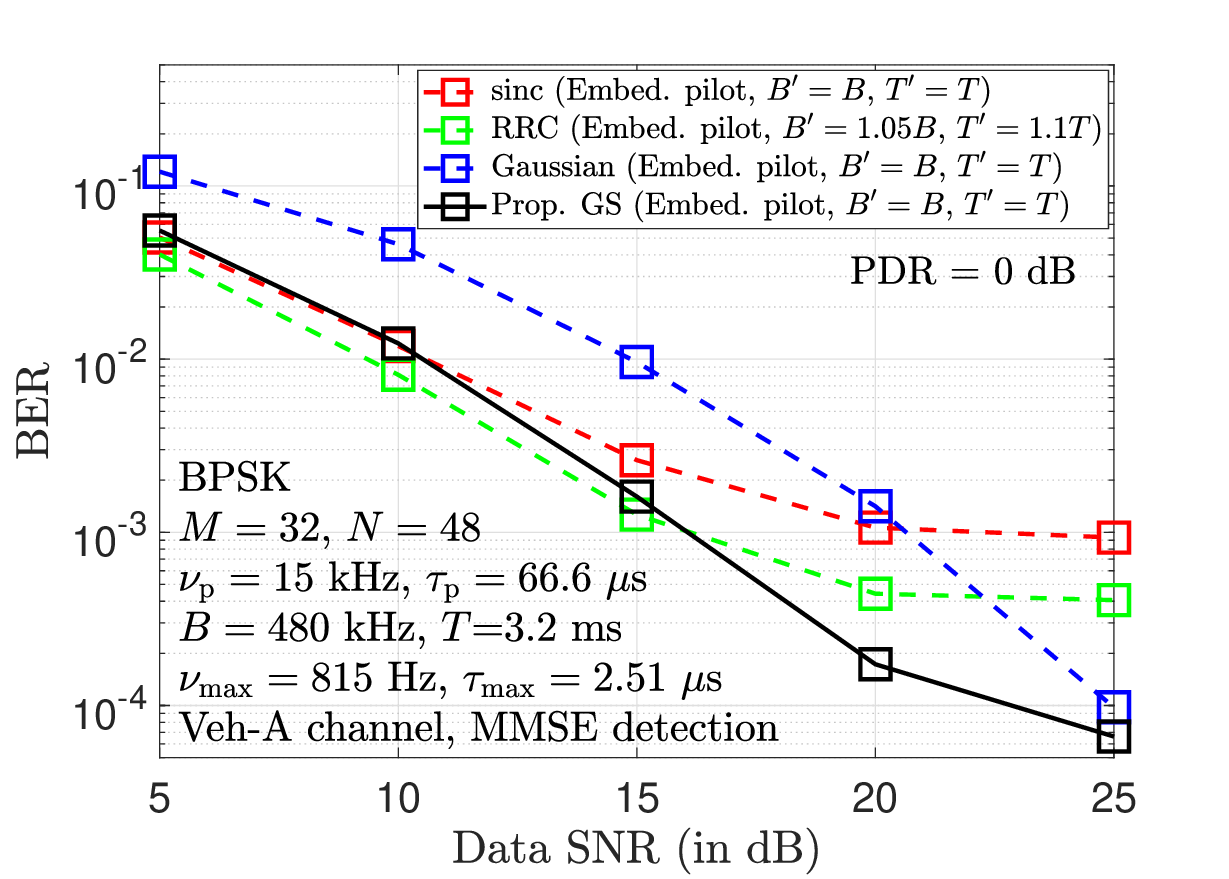}
\caption{BER vs data SNR performance of different filters with embedded pilot frame at 0 dB PDR for BPSK.}
\label{fig_ber}
\vspace{-5mm}
\end{figure}

\begin{figure}[!t]
\centering
\includegraphics[width=9.0cm,height=6.5cm]{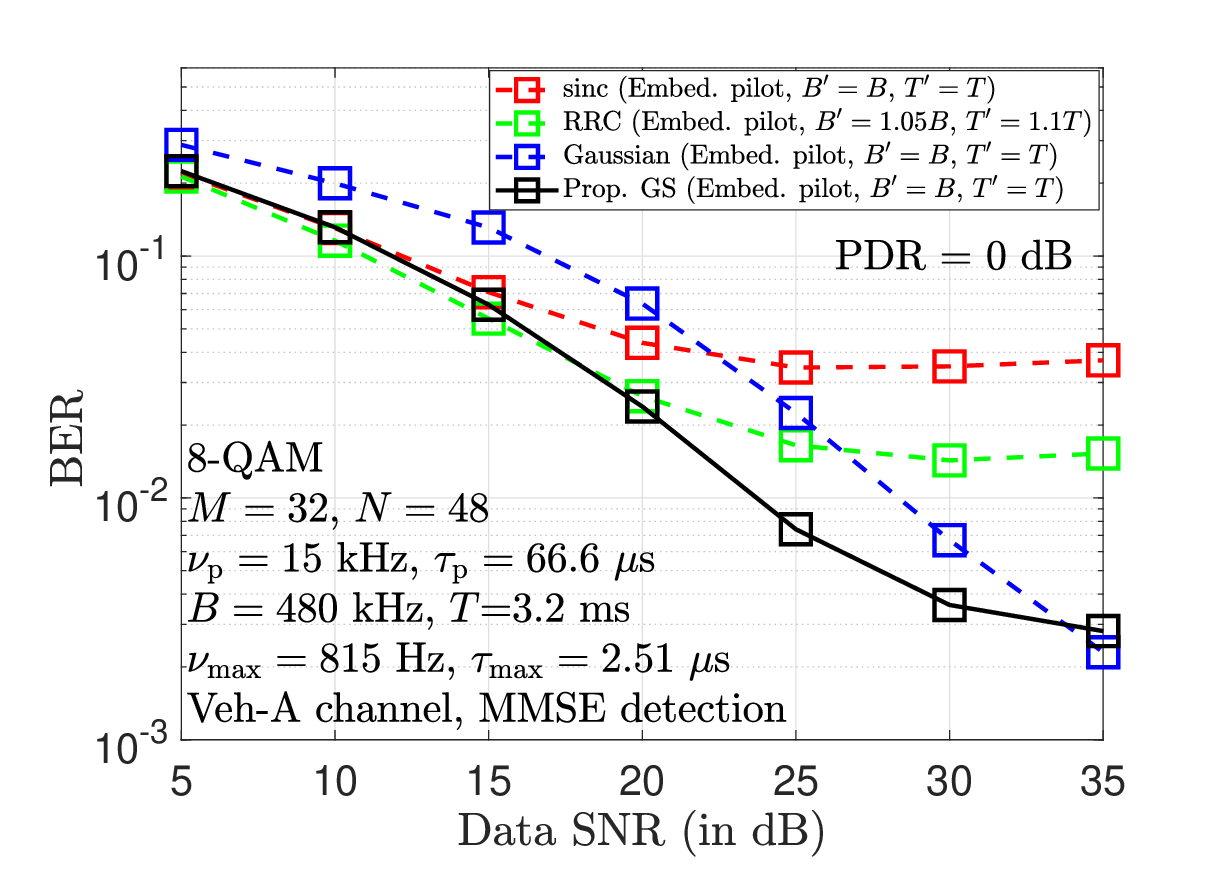}
\caption{BER vs data SNR performance of different filters with embedded pilot frame at 0 dB PDR for 8-QAM.}
\label{fig_ber_8qam}
\vspace{-5mm}
\end{figure}

In Fig. \ref{fig_ber}, we present the BER performance of different filters corresponding to the MSE performance depicted in Fig. \ref{fig_mse}. At low SNRs, the Gaussian filter exhibits the poorest performance among the evaluated filters. This can be attributed to its relatively high main lobe value at the sampling instants, significantly increasing inter-symbol interference and consequently degrading data detection performance when the noise effects are more pronounced. While the sinc filter outperforms the Gaussian filter at low SNRs, its significant side lobe levels cause substantial pilot-data interference. As a result, the BER performance of the sinc filter starts to floor around 20 dB data SNR, as evident from the corresponding MSE performance. The proposed GS filter consistently outperforms the sinc filter with flooring beyond 25 dB. This improved performance is attributed to its lower side lobe levels compared to the sinc filter, which reduces interference, consequently improving both channel estimation and data detection accuracy. The RRC filter with bandwidth/time expansion, also due to its favorable side lobe properties, outperforms sinc filter but performs poorer than the proposed GS filter. However, the Gaussian filter has crossovers with these non-Gaussian filters. Notably, these crossover points in BER performance closely align with the SNR values where the MSE performance of the filters begin to floor as shown in Fig. \ref{fig_mse}. As the MSE of non-Gaussian filters floors, further SNR improvements do not significantly enhance channel estimation accuracy, which ultimately impacts the detection performance.

\begin{figure}[!t]
\centering
\includegraphics[width=9.0cm,height=6.5cm]{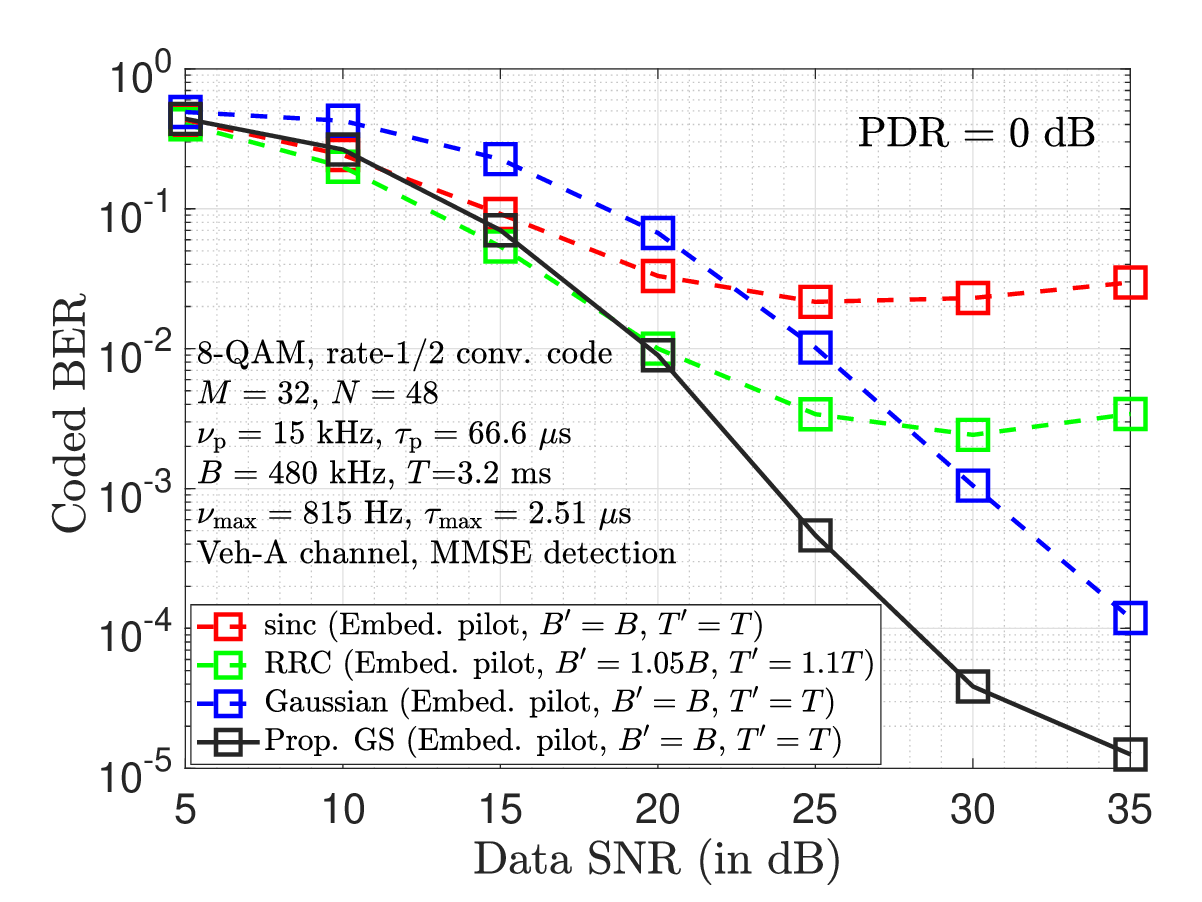}
\caption{Coded BER vs data SNR performance of different filters with embedded pilot frame at 0 dB PDR for 8-QAM and rate-1/2 coding.}
\label{fig_ber_8qam_coded}
\vspace{-7mm}
\end{figure}

Figure \ref{fig_ber_8qam} shows the BER performance of different filters with 8-QAM and embedded pilot frame at a PDR of 0 dB. It can be seen that the performance of the proposed GS filter is superior compared to Gaussian and sinc filters. For example, at a BER of $10^{-2}$, the GS filter achieves an SNR gain of about 4 dB compared to Gaussian and sinc filters. Corresponding to the uncoded BER performance in Fig. \ref{fig_ber_8qam}, Fig. \ref{fig_ber_8qam_coded} shows the coded BER performance with a rate-$1/2$ convolutional code with constraint length 7. From Fig. \ref{fig_ber_8qam_coded}, we can see that the proposed GS filter achieves an SNR gain in excess of 6 dB at a coded BER of $10^{-4}$ compared to Gaussian and sinc filters.

\vspace{-2mm}
\section{Conclusion}
\label{sec6}
In this paper, we brought out the role of DD domain pulse shaping filters on the individual performance of model-free I/O relation estimation and equalization/detection in Zak-OTFS. The Gaussian and sinc filters reported in the literature were shown to possess complementary strengths with respect to these two receiver tasks, viz., Gaussian pulse shape is good for the I/O relation estimation task but poor for the equalization/detection task, whereas sinc pulse shape is poor for I/O relation estimation but good for equalization/detection. Based on this observation, we proposed a new filter, termed as Gaussian-sinc (GS) filter, which inherited the complementary strengths of both Gaussian and sinc filters. The proposed filter did not incur time or bandwidth expansion. We derived closed-form expressions for the I/O relation and noise covariance of Zak-OTFS with the proposed GS filter. Our simulation results with Veh-A fractional DD channels using model-free I/O relation estimation with exclusive and embedded pilot frames showed that the proposed GS filter achieves better BER performance compared to Gaussian and sinc filters (e.g., an SNR gain in excess of 6 dB at a coded BER of $10^{-4}$ in favor of the proposed filter). The GS filter performance in comparison with those of other filters using superimposed/spread pilot (where there will be no pilot/guard regions and consequent loss in throughput) can be investigated as future work.

\vspace{-2mm}
\appendices
\section{Energy normalization expressions for GS filter}
\label{appxA}
The delay domain filter in (\ref{delay_domain}) should have unit normalized energy, i.e.,
\begin{align}
\int |w_1(\tau)|^{2}d\tau=\Omega_{\tau}^{2}B\int \mathrm{sinc}^{2}(B\tau)e^{-2\alpha_{\tau}B^{2}\tau^{2}}d\tau=1.
\label{normalize}
\end{align}
Assume $x_1(\tau)=\mathrm{sinc}^{2}(B\tau)$ and $x_2^{*}(\tau)=e^{-2\alpha_{\tau}B^{2}\tau^{2}}$. The frequency domain representation of the delay domain signals $x_1(\tau)$ and $x_2(\tau)$, denoted by $X_1(f)$ and $X_2(f)$, respectively, 
are given by
\begin{align}
X_1(f)=\left(\frac{B-|f|}{B^{2}}\right)\mathbbm{1}_{\{0\leq |f|\leq B\}},
\end{align}
\begin{align}
X_2(f)=\sqrt{\frac{\pi}{2\alpha_{\tau}B^{2}}}e^{-\frac{\pi^{2}f^{2}}{2\alpha_{\tau}B^{2}}}.
\end{align}
Using Parseval's theorem, the integral in (\ref{normalize}) can be written as
\vspace{0mm}
\begin{eqnarray}
\int |w_{1}(\tau)|^{2}d\tau & \hspace{-2mm} = & \hspace{-2mm} \left(\frac{\Omega_{\tau}^{2}}{B}\right)\sqrt{\frac{\pi}{2\alpha_{\tau}}}\int X_1(f)X_2^{*}(f)df \nonumber \\
& \hspace{-40mm} = & \hspace{-21mm} \left(\frac{\Omega_{\tau}^{2}}{B^{2}}\right)\sqrt{\frac{\pi}{2\alpha_{\tau}}}\Bigg(B\int_{-B}^{B}e^{-\frac{\pi^{2}f^{2}}{2\alpha_{\tau}B^{2}}}df-\int_{0}^{B}fe^{-\frac{\pi^{2}f^{2}}{2\alpha_{\tau}B^{2}}}df \nonumber \\ 
& \hspace{-40mm} & \hspace{-22mm} +\int_{-B}^{0}fe^{-\frac{\pi^{2}f^{2}}{2\alpha_{\tau}B^{2}}}df \Bigg) \nonumber \\
& \hspace{-40mm} = & \hspace{-21mm} \Omega_{\tau}^{2}\sqrt{\frac{\pi}{2\alpha_{\tau}}}\left(\sqrt{\frac{2\alpha_{\tau}}{\pi}}\mathrm{erf}\left(\frac{\pi}{\sqrt{2\alpha_{\tau}}}\right)-\frac{2\alpha_{\tau}}{\pi^{2}}\left(1-e^{-\frac{\pi^{2}}{2\alpha_{\tau}}}\right)\right) \nonumber \\
& \hspace{-40mm} = & \hspace{-21mm} \Omega_{\tau}^{2}\left(\mathrm{erf}\left(\frac{\pi}{\sqrt{2\alpha_{\tau}}}\right)-\sqrt{\frac{2\alpha_{\tau}}{\pi^{3}}}\left(1-e^{-\frac{\pi^{2}}{2\alpha_{\tau}}}\right)\right).
\label{normalize_1}
\end{eqnarray}

Equating (\ref{normalize_1}) to $1$, we get
\begin{align}
\Omega_{\tau}=\frac{1}{\sqrt{\left(\mathrm{erf}\left(\frac{\pi}{\sqrt{2\alpha_{\tau}}}\right)-\sqrt{\frac{2\alpha_{\tau}}{\pi^{3}}}\left(1-e^{-\frac{\pi^{2}}{2\alpha_{\tau}}}\right)\right)}}.
\label{Omega_tau}
\end{align}
A similar line of derivation for the Doppler domain filter in (\ref{Doppler_domain}) gives 
\begin{align}
\Omega_{\nu}=\frac{1}{\sqrt{\left(\mathrm{erf}\left(\frac{\pi}{\sqrt{2\alpha_{\nu}}}\right)-\sqrt{\frac{2\alpha_{\nu}}{\pi^{3}}}\left(1-e^{-\frac{\pi^{2}}{2\alpha_{\nu}}}\right)\right)}}.
\label{Omega_nu}
\end{align}

\section{Derivation of (\ref{eqn:GS_match_channel})}
\label{appxB}
First, consider the $I_{i}^{(1)}(\tau)$ term in (\ref{eqn:channel_matched}). 
Using the $w_1({\tau})$ expression for the GS filter in (\ref{delay_domain}), the  $I_{i}^{(1)}(\tau)$ term in (\ref{eqn:channel_matched}) can be written as
\begin{eqnarray}
\hspace{-8mm}
I_{i}^{(1)}(\tau) & \hspace{-2mm} = & \hspace{-2mm}  \Omega_{\tau}^{2}B\int \underbrace{e^{-\alpha_{\tau}B^{2}\tau_{1}^{2}}e^{-\alpha_{\tau}B^{2}(\tau-\tau_{i}-\tau_{1})^{2}}e^{-j2\pi\nu_{i}\tau_{1}}}_{\overset{\Delta}{=}x_{1}(\tau_{1})} \nonumber \\
& \hspace{-2mm} & \hspace{-2mm} \underbrace{\mathrm{sinc}(B\tau_{1})\mathrm{sinc}(B(\tau-\tau_{i}-\tau_{1}))}_{\overset{\Delta}{=}x_{2}^{*}(\tau_{1})}d\tau_{1} \nonumber \\
& \hspace{-20mm} = & \hspace{-11mm} \Omega_{\tau}^{2}B\int X_{1}(f)X_{2}^{*}(f)df, \ \ \ \text{ (by Parseval's theorem)}
\label{integral_1}
\end{eqnarray}
where $X_{1}(f)$ and $X_{2}(f)$ are the frequency domain representations of the delay domain signals $x_{1}(\tau_{1})$ and $x_{2}(\tau_{1})$, respectively, given by
\begin{equation}
\hspace{-4mm}
X_{1}(f)= \kappa_\tau 
e^{-\frac{\alpha_{\tau}B^{2}}{2}\left((\tau-\tau_i)^{2}+2j\frac{\pi(f+\nu_{i})(\tau-\tau_i)}{\alpha_{\tau}B^{2}}+\frac{\pi^{2}(f+\nu_{i})^{2}}{(\alpha_{\tau}B^{2})^{2}}\right)},
\end{equation}
where $\kappa_\tau = \sqrt{\frac{\pi}{2\alpha_{\tau}B^{2}}}$, and
\begin{eqnarray}
X_{2}(f) & \hspace{-2mm}  = & \hspace{-2mm} \frac{1}{B^{2}} \Bigg[\frac{1}{j2\pi(\tau-\tau_{i})}\bigg(\Big(e^{j\pi (B-2f)(\tau-\tau_{i})} \nonumber \\
& \hspace{-25mm} & \hspace{-17mm} 
-e^{-j\pi B(\tau-\tau_{i})}\Big) 
\mathbbm{1}_{\{0<f<B\}}  \hspace{-1mm} +  \hspace{-1mm} \left(e^{j\pi B(\tau-\tau_{i})}-e^{-j\pi (2f+B)(\tau-\tau_{i})}\right) \nonumber \\
& \hspace{-25mm} & \hspace{-17mm} 
\mathbbm{1}_{\{-B<f<0\}}\bigg)\Bigg] 
\mathbbm{1}_{\{\tau\neq\tau_{i}\}}\hspace{-0mm}+\hspace{-0mm}\frac{1}{B^{2}}\Big[(B-f)\mathbbm{1}_{\{0<f<B\}} 
\nonumber \\
& \hspace{-25mm} & \hspace{-17mm}
+(B+f)\mathbbm{1}_{\{-B<f<0\}}\Big]\mathbbm{1}_{\{\tau=\tau_{i}\}}.
\end{eqnarray}

Now, consider the case of $\tau \neq \tau_i$.
Substituting $z=f+\nu_{i}$, (\ref{integral_1}) becomes
\begin{eqnarray}
I_{i,1}^{(1)}(\tau) & \hspace{-2.5mm} = & \hspace{-2.5mm} C_{1,1}(\tau,\tau_i)\Bigg[e^{j\pi B(\tau-\tau_{i})}\hspace{-1.5mm}\underbrace{\int_{\nu_{i}}^{B+\nu_{i}}\hspace{-4mm}e^{-\frac{\pi^{2}}{2\alpha_{\tau}B^{2}}z^{2}-j\pi(\tau-\tau_i) z}dz}_{\overset{\Delta}{=}I_{i,1,1}^{(1)}(\tau)} \nonumber \\
& \hspace{-15mm} & \hspace{-10mm} 
-e^{-j\pi (B+2\nu_{i})(\tau-\tau_{i})}\underbrace{\int_{\nu_{i}}^{B+\nu_{i}}e^{-\frac{\pi^{2}}{2\alpha_{\tau}B^{2}}z^{2}+j\pi (\tau-\tau_i)z}dz}_{\overset{\Delta}{=}I_{i,1,2}^{(1)}(\tau)} \nonumber \\
& \hspace{-15mm} & \hspace{-10mm} 
+e^{j\pi (B-2\nu_{i})(\tau-\tau_{i})}\underbrace{\int_{-B+\nu_{i}}^{\nu_{i}}e^{-\frac{\pi^{2}}{2\alpha_{\tau}B^{2}}z^{2}+j\pi (\tau-\tau_i)z}dz}_{\overset{\Delta}{=}I_{i,1,3}^{(1)}(\tau)} \nonumber \\
& \hspace{-15mm} & \hspace{-10mm} -e^{-j\pi B(\tau-\tau_{i})}\underbrace{\int_{-B+\nu_{i}}^{\nu_{i}}e^{-\frac{\pi^{2}}{2\alpha_{\tau}B^{2}}z^{2}-j\pi (\tau-\tau_i)z}dz}_{\overset{\Delta}{=}I_{i,1,4}^{(1)}(\tau)}\Bigg], 
\label{integral_1_1}
\end{eqnarray}
where $C_{1,1}(\tau,\tau_i)\overset{\Delta}{=}\Omega_{\tau}^{2}\sqrt{\frac{\pi}{2\alpha_{\tau}}}\left(\frac{e^{-\frac{\alpha_{\tau}B^{2}}{2}(\tau-\tau_i)^{2}}}{j2\pi B^{2}(\tau-\tau_i)}\right)$, 
\begin{eqnarray}
f_{1}(a,x,y,z) & \hspace{-2mm}  \overset{\Delta}{=} & \hspace{-2mm}
\frac{e^{-\frac{z^{2}}{4a}}\sqrt{\pi}}{2\sqrt{a}} \Big[\mathrm{erf}\left(\sqrt{a}\left(x+\frac{jz}{2a}\right)\right) \nonumber \\
& \hspace{-2mm} & \hspace{-2mm} 
-\mathrm{erf}\left(\sqrt{a}\left(y+\frac{jz}{2a}\right)\right)\Big], 
\label{f_1}
\end{eqnarray}
and the integral terms in ($\ref{integral_1_1}$) become \\
$I_{i,1,1}^{(1)}(\tau)= f_{1}\left(a_1,B+\nu_i,\nu_i,b_i^{(1)}\right)$, \\
$I_{i,1,2}^{(1)}(\tau)= f_{1}\left(a_1,B+\nu_i,\nu_i,-b_i^{(1)}\right)$, \\
$I_{i,1,3}^{(1)}(\tau) = f_{1}\left(a_1,\nu_i,\nu_i-B,-b_i^{(1)}\right)$, \\
$I_{i,1,4}^{(1)}(\tau) = f_{1}\left(a_1,\nu_i,\nu_i-B,b_i^{(1)}\right)$,
where $a_1=\frac{\pi^{2}}{2\alpha_{\tau}B^{2}}$ and 
$b_{i}^{(1)}\overset{\Delta}{=}\pi(\tau-\tau_i)$. 

Now, consider the case of $\tau=\tau_i$.
Substituting $z=f+\nu_{i}$, the integral in (\ref{integral_1}) becomes
\begin{eqnarray}
I_{i,2}^{(1)}(\tau) & \hspace{-2mm} = & \hspace{-2mm} C_{1,2}\Bigg[(B+\nu_i)\underbrace{\int_{\nu_{i}}^{B+\nu_{i}}e^{-\frac{\pi^{2}}{2\alpha_{\tau}B^{2}}z^{2}}dz}_{\overset{\Delta}{=}I_{i,2,1}^{(1)}} \nonumber \\
& \hspace{-25mm} & \hspace{-17mm} -\underbrace{\int_{\nu_{i}}^{B+\nu_{i}}ze^{-\frac{\pi^{2}}{2\alpha_{\tau}B^{2}}z^{2}}dz}_{\overset{\Delta}{=}I_{i,2,2}^{(1)}} +(B-\nu_i)\underbrace{\int_{-B+\nu_{i}}^{\nu_{i}}e^{-\frac{\pi^{2}}{2\alpha_{\tau}B^{2}}z^{2}}dz}_{\overset{\Delta}{=}I_{i,2,3}^{(1)}} \nonumber \\
& \hspace{-25mm} & \hspace{-17mm} +\underbrace{\int_{-B+\nu_{i}}^{\nu_{i}}ze^{-\frac{\pi^{2}}{2\alpha_{\tau}B^{2}}z^{2}}dz}_{\overset{\Delta}{=}I_{i,2,4}^{(1)}}\Bigg], 
\label{integral_1_2}
\end{eqnarray}
where 
$f_2(a,x,y)\overset{\Delta}{=}\frac{\sqrt{\pi}}{2\sqrt{a}}\left(\mathrm{erf}(\sqrt{a}x)-\mathrm{erf}(\sqrt{a}y)\right)$, 
$C_{1,2}=\Omega_{\tau}^{2}\sqrt{\frac{\pi}{2\alpha_{\tau}}}\left(\frac{1}{B^{2}}\right)$, 
$f_3(a,x,y)\overset{\Delta}{=}\frac{1}{2a}\left(e^{-a x^{2}}-e^{-a y^{2}}\right)$,
and the integrals in (\ref{integral_1_2}) become
$I_{i,2,1}^{(1)}=f_2(a_1,B+\nu_i,\nu_i)$,
$I_{i,2,2}^{(1)}=f_3(a_1,\nu_i,B+\nu_i)$,
$I_{i,2,3}^{(1)}=f_2(a_1,\nu_i,\nu_i-B)$,
and
$I_{i,2,4}^{(1)}=f_3(a_1,\nu_i-B,\nu_i)$.

Next, consider the $I_{i}^{(2)}(\tau,\nu)$ term in (\ref{eqn:channel_matched}). For the GS filter in (\ref{delay_domain}), the  $I_{i}^{(2)}(\tau,\nu)$ term in (\ref{eqn:channel_matched}) can be written as
\begin{eqnarray}
I_{i}^{(2)}(\tau,\nu) & \hspace{-2mm} = & \hspace{-2mm} \Omega_{\nu}^{2}T\int \underbrace{e^{-\alpha_{\nu}T^{2}\nu_{1}^{2}}e^{-\alpha_{\nu}T^{2}(\nu-\nu_{i}-\nu_{1})^{2}}e^{+j2\pi\nu_{1}\tau}}_{\overset{\Delta}{=}X_{1}(\nu_{1})} \nonumber \\
& \hspace{-2mm} & \hspace{-2mm} \underbrace{\mathrm{sinc}(T\nu_{1})\mathrm{sinc}(T(\nu-\nu_{i}-\nu_{1}))}_{\overset{\Delta}{=}X_{2}^{*}(\nu_{1})}d\tau_{1} \nonumber \\
& \hspace{-27mm} = & \hspace{-15mm} \Omega_{\nu}^{2}T
\int x_{1}(t)x_{2}^{*}(t)dt, \quad \text{ (by Parseval's theorem)}
\label{integral_2}
\end{eqnarray}
where $x_1(t)$ and $x_2(t)$ are the time domain representations of Doppler domain signals $X_1(\nu_1)$ and $X_2(\nu_1)$, respectively, given by
\begin{equation}
x_1(t)=\kappa_\nu
e^{-\frac{\alpha_{\nu}T^{2}}{2}\left((\nu-\nu_i)^{2}-2j\frac{\pi(t+\tau)(\nu-\nu_i)}{\alpha_{\nu}T^{2}}+\frac{\pi^{2}(t+\tau)^{2}}{(\alpha_{\nu}T^{2})^{2}}\right)}, 
\end{equation}
where $\kappa_\nu = \sqrt{\frac{\pi}{2\alpha_{\nu}T^{2}}}$, and 
\begin{eqnarray}
x_{2}(t) & \hspace{-2mm} = & \hspace{-2mm} \frac{1}{j2\pi T^{2}(\nu-\nu_i)} \Bigg[\bigg(\left(e^{j\pi (\nu-\nu_i)T}-e^{j\pi (\nu-\nu_i)(2t-T)}\right) \nonumber \\
& \hspace{-2mm} & \hspace{-2mm}
\mathbbm{1}_{\{0<t<T\}} + \left(e^{j\pi (\nu-\nu_i)(2t+T)}-e^{-j\pi (\nu-\nu_i)T}\right)
\nonumber \\
& \hspace{-2mm} & \hspace{-2mm} 
\mathbbm{1}_{\{-T<t<0\}}\bigg)\Bigg] 
\mathbbm{1}_{\{\nu\neq\nu_{i}\}}+\frac{1}{T^{2}}\Big[(T-t)\mathbbm{1}_{\{0<t<T\}}
\nonumber \\
& \hspace{-2mm} & \hspace{-2mm} 
+(T+t)\mathbbm{1}_{\{-T<t<0\}}\Big]\mathbbm{1}_{\{\nu=\nu_{i}\}}.
\end{eqnarray}
For the case of $\nu\neq\nu_i$, substituting $z=t+\tau$, (\ref{integral_2}) becomes
\begin{eqnarray}
I_{i,1}^{(2)}(\tau,\nu) & \hspace{-2mm} = & \hspace{-2mm} C_{2,1}(\nu,\nu_i)\Bigg[e^{j\pi (\nu-\nu_{i})(T+2\tau)} 
\underbrace{\hspace{-1mm}\int_{\tau}^{\tau+T}\hspace{-3mm}e^{-\frac{\pi^{2}z^{2}}{2\alpha_{\nu}T^{2}}-j\pi(\nu-\nu_i)z}\hspace{-1mm}dz}_{\overset{\Delta}{=}I_{i,1,1}^{(2)}(\tau,\nu)} 
\nonumber \\
& \hspace{-2mm} & \hspace{-2mm} 
-e^{-j\pi(\nu-\nu_i)T}
\underbrace{\hspace{-1mm}\int_{\tau}^{T+\tau}\hspace{-2mm}e^{-\frac{\pi^{2}z^{2}}{2\alpha_{\nu}T^{2}}+j\pi(\nu-\nu_i)z}dz}_{\overset{\Delta}{=}I_{i,1,2}^{(2)}(\tau,\nu)} 
\nonumber \\
& \hspace{-2mm} & \hspace{-2mm}  
+e^{j\pi(\nu-\nu_i)T}
\underbrace{\hspace{-1mm}\int_{-T+\tau}^{\tau}\hspace{-2mm}e^{-\frac{\pi^{2}z^{2}}{2\alpha_{\nu}T^{2}}+j\pi(\nu-\nu_i)z}dz}_{\overset{\Delta}{=}I_{i,1,3}^{(2)}(\tau,\nu)} 
\nonumber \\
& \hspace{-2mm} & \hspace{-2mm} 
-e^{-j\pi(\nu-\nu_i)(T-2\tau)}
\underbrace{\hspace{-1mm}\int_{-T+\tau}^{\tau}\hspace{-2mm} e^{-\frac{\pi^{2}z^{2}}{2\alpha_{\nu}T^{2}}-j\pi(\nu-\nu_i)z}dz}_{\overset{\Delta}{=}I_{i,1,4}^{(2)}(\tau,\nu)} \Bigg],
\label{integral_2_1}
\end{eqnarray}
where $C_{2,1}(\nu,\nu_i)=\Omega_{\nu}^{2}\sqrt{\frac{\pi}{2\alpha_{\nu}}}\left(\frac{e^{-\frac{\alpha_{\nu}T^{2}}{2}(\nu-\nu_i)^{2}}}{j2\pi T^{2}(\nu-\nu_i)}\right)$, 
and the integral terms in (\ref{integral_2_1}) become \\
$I_{i,1,1}^{(2)}(\tau,\nu)=f_1\left(a_2,\tau+T,\tau,b_i^{(2)}\right)$, \\
$I_{i,1,2}^{(2)}(\tau,\nu)=f_1\left(a_2,\tau+T,\tau,-b_i^{(2)}\right)$, \\
$I_{i,1,3}^{(2)}(\tau,\nu)=f_1\left(a_2,\tau,\tau-T,-b_i^{(2)}\right)$, \\
$I_{i,1,4}^{(2)}(\tau,\nu)=f_1\left(a_2,\tau,\tau-T,b_i^{(2)}\right)$,
where $a_2=\frac{\pi^{2}}{2\alpha_{\nu}T^{2}}$ and
$b_i^{(2)}\overset{\Delta}{=}\pi(\nu-\nu_i)$. 

For the case of $\nu=\nu_i$, substituting $z=t+\tau$, the integral in (\ref{integral_2}) becomes
\begin{eqnarray}
I_{i,2}^{(2)}(\tau,\nu) & \hspace{-2mm} = & \hspace{-2mm} C_{2,2}\Bigg[(T+\tau)\underbrace{\int_{\tau}^{T+\tau}e^{-\frac{\pi^{2}}{2\alpha_{\nu}T^{2}}z^{2}}dz}_{\overset{\Delta}{=}I_{i,2,1}^{(2)}(\tau)} \nonumber \\
& \hspace{-20mm} & \hspace{-15mm} -\underbrace{\int_{\tau}^{T+\tau}ze^{-\frac{\pi^{2}}{2\alpha_{\nu}T^{2}}z^{2}}dz}_{\overset{\Delta}{=}I_{i,2,2}^{(2)}(\tau)}+(T-\tau)\underbrace{\int_{-T+\tau}^{\tau}e^{-\frac{\pi^{2}}{2\alpha_{\nu}T^{2}}z^{2}}dz}_{\overset{\Delta}{=}I_{i,2,3}^{(2)}(\tau)} \nonumber \\
& \hspace{-20mm} & \hspace{-15mm}
+\underbrace{\int_{-T+\tau}^{\tau}ze^{-\frac{\pi^{2}}{2\alpha_{\nu}T^{2}}z^{2}}dz}_{\overset{\Delta}{=}I_{i,2,4}^{(2)}(\tau)}\Bigg], 
\label{integral_2_2}
\end{eqnarray}
where $C_{2,2}=\Omega_{\nu}^{2}\sqrt{\frac{\pi}{2\alpha_{\nu}}}\left(\frac{1}{T^{2}}\right)$, and the integrals in (\ref{integral_2_2}) 
become
$I_{i,2,1}^{(2)}(\tau)=f_2(a_2,\tau+T,\tau)$,
$I_{i,2,2}^{(2)}(\tau)=f_3(a_2,\tau,\tau+T)$,
$I_{i,2,3}^{(2)}(\tau)=f_2(a_2,\tau,\tau-T)$,
$I_{i,2,4}^{(2)}(\tau)=f_3(a_2,\tau-T,\tau)$. 

Combining the expressions for $I_{i,1}^{(1)}(\tau)$ in (\ref{integral_1_1}) for $\tau\neq \tau_i$ and $I_{i,2}^{(1)}(\tau)$ in (\ref{integral_1_2}) for $\tau=\tau_i$, we get $I_{i}^{(1)}(\tau)$ as
$I_{i}^{(1)}(\tau)=I_{i,1}^{(1)}(\tau)\mathbbm{1}_{\{\tau\neq\tau_i\}}+I_{i,2}^{(1)}(\tau)\mathbbm{1}_{\{\tau=\tau_i\}}$.
Similarly, combining the expressions for $I_{i,1}^{(2)}(\tau,\nu)$ in (\ref{integral_2_1}) for $\nu\neq \nu_i$ and $I_{i,2}^{(2)}(\tau,\nu)$ in (\ref{integral_2_2}) for $\nu=\nu_i$, 
we get $I_{i}^{(2)}(\tau, \nu)$ as
$I_{i}^{(2)}(\tau, \nu)=I_{i,1}^{(2)}(\tau,\nu)\mathbbm{1}_{\{\nu\neq\nu_i\}}+I_{i,1}^{(2)}(\tau,\nu)\mathbbm{1}_{\{\nu=\nu_i\}}$. These $I_{i}^{(1)}(\tau)$ and $I_{i}^{(2)}(\tau, \nu)$ expressions used in (\ref{eqn:channel_matched}) gives the effective channel expression in (\ref{eqn:GS_match_channel}).  

The erf(.) functions in the derived expressions can be computed using accurate closed-form approximations for the erf(.) function \cite{erf1},\cite{erf2}.

\section{Derivation of (\ref{eqn:noise_expectation})}
\label{appxC}
For the GS filter in (\ref{delay_domain}),
the term $I_q^{(3)}(\tau)$ in (\ref{eqn:noise_matched}) can be written as
\begin{eqnarray}
\hspace{-8mm}
I_q^{(3)}(\tau) & \hspace{-2mm} = & \hspace{-2mm} \Omega_{\nu}\sqrt{T}\int \underbrace{e^{-\alpha_{\nu}T^{2}\nu_1^{2}}e^{j2\pi\nu_1(\tau+q\tau_p)}}_{\overset{\Delta}{=}X_1(\nu_1)}\underbrace{\mathrm{sinc}(T\nu_1)}_{\overset{\Delta}{=}X_2^{*}(\nu_1)}d\nu_1 \nonumber \\
& \hspace{-17mm} = & \hspace{-10mm} \Omega_{\nu}\sqrt{T}\int x_1(t)x_2^{*}(t)dt, \quad \text{ (by Parseval's theorem)}
\label{integral_3}
\end{eqnarray} 
where $x_1(t)$ and $x_2(t)$ are the time domain representations of $X_1(\nu_1)$ and $X_2(\nu_1)$, respectively, given by
\begin{equation}
x_1(t)=\sqrt{\frac{\pi}{\alpha_{\nu}T^{2}}}e^{-\frac{\pi^{2}(\tau+q\tau_p+t)^{2}}{\alpha_{\nu}T^{2}}},
\label{noise_1}
\end{equation}
\begin{equation}
x_2(t)=\frac{1}{T}\mathrm{rect}\left(\frac{t}{T}\right). 
\label{noise_2}
\end{equation}
Carrying out the integration in (\ref{integral_3}) using (\ref{noise_1}) and (\ref{noise_2}) gives
\begin{eqnarray}
I_{q}^{(3)}(\tau) & \hspace{-2mm} = & \hspace{-2mm} \frac{\Omega_{\nu}}{T^{\frac{3}{2}}}\sqrt{\frac{\pi}{\alpha_{\nu}}}\int_{-\frac{T}{2}}^{\frac{T}{2}}e^{-\frac{\pi^{2}(\tau+q\tau_p+t)^{2}}{\alpha_{\nu}T^{2}}}dt \nonumber \\
& \hspace{-2mm} = & \hspace{-2mm} \frac{\Omega_{\nu}}{2\sqrt{T}}\Bigg(\mathrm{erf}\left(\sqrt{\frac{\pi^{2}}{\alpha_{\nu}T^{2}}}\left(\tau+q\tau_p+\frac{T}{2}\right)\right) \nonumber \\
& \hspace{-2mm} & \hspace{-2mm} -\mathrm{erf}\left(\sqrt{\frac{\pi^{2}}{\alpha_{\nu}T^{2}}}\left(\tau+q\tau_p-\frac{T}{2}\right)\right)\Bigg) \nonumber \\
& \hspace{-2mm} \overset{\Delta}{=} & \hspace{-2mm} g(\tau+q\tau_p).
\end{eqnarray}
Sampling (\ref{eqn:noise_matched}) on $\Lambda_{\mathrm{dd}}$ gives
\begin{align}
n_{\mathrm{dd}}[k,l]=&\sqrt{\tau_p}\sum_{q=-\infty}^{\infty}e^{-j2\pi\frac{ql}{N}}g\left(\frac{k\tau_p}{M}+q\tau_p\right) \nonumber \\
&\left(\int w_1^{*}(-\tau_1)n\left(\frac{k\tau_p}{M}-\tau_1+q\tau_p\right)d\tau_1\right).
\end{align}
The $(k_{1}N+l_1+1,k_{2}N+l_{2}+1)$th term of the noise covariance matrix is given by 
\begin{align}
&\mathbbm{E}[n_{\mathrm{dd}}[k_1,l_1],n_{\mathrm{dd}}^{*}[k_2,l_2]]=\tau_p\sum_{q_1=-\infty}^{\infty}\sum_{q_2=-\infty}^{\infty}e^{j2\pi\frac{q_2l_2-q_1l_1}{N}} \nonumber \\
&g\left(\frac{k_1\tau_p}{M}+q_1\tau_p\right)g^{*}\left(\frac{k_2\tau_p}{M}+q_2\tau_p\right)\bigg(\iint w_1^{*}(-\tau_1)w_1^{*}(-\tau_2) \nonumber \\
&\underbrace{\mathbbm{E}\bigg[n\left(\frac{k_1\tau_p}{M}-\tau_1+q_1\tau_p\right)n^{*}\left(\frac{k_2\tau_p}{M}-\tau_2+q_2\tau_p\right)\bigg]}_{=N_{0}\delta \left(\tau_2-\tau_1-\left(\frac{k_2-k_1}{M}\right)\tau_p-(q_2-q_1)\tau_p \right)}d\tau_1 d\tau_2\bigg).
\label{integral_3_1}
\end{align}
Defining the term 
$x_{\{k_1,k_2,q_1,q_2\}}\overset{\Delta}{=}\left(\frac{k_2-k_1}{M}\right)\tau_p+(q_2-q_1)\tau_p$, the double integral in (\ref{integral_3_1}) becomes
\begin{eqnarray}
S_{\{k_1,k_2,q_1,q_2\}} & \hspace{-3.5mm} \overset{\Delta}{=} & \hspace{-2.5mm} \Omega_{\tau}^{2}N_{0}B\Bigg[\hspace{-1mm}\int \underbrace{\hspace{-2mm} \mathrm{sinc}(B\tau_1)\mathrm{sinc}\Big(\hspace{-1.0mm}B\hspace{-0.25mm}\big(\hspace{-0.25mm}\tau_1+x_{\{k_1,k_2,q_1,q_2\}}\big)\hspace{-0.5mm}\Big)}_{\overset{\Delta}{=}x_{2}^{*}(\tau_1)} \nonumber \\
& \hspace{-3.5mm} & \hspace{-2.5mm} \underbrace{e^{-\alpha_{\tau}B^{2}\tau_1^{2}}e^{-\alpha_{\tau}B^{2}\left(\tau_1+x_{\{k_1,k_2,q_1,q_2\}}\right)^{2}}}_{\overset{\Delta}{=}x_1(\tau_1)}d\tau_1\Bigg] \nonumber \\
& \hspace{-3.5mm} = & \hspace{-2.5mm} \Omega_{\tau}^{2}N_{0}B\int X_{2}^{*}(f)X_1(f)df,
\label{integral_3_2}
\end{eqnarray}
where $X_{1}(f)$ and $X_{2}(f)$ are the frequency domain representations of $x_1(\tau_1)$ and $x_2(\tau_1)$, respectively, given by
\begin{align}
X_{1}(f)=\kappa_\tau
e^{-\frac{\alpha_{\tau}B^{2}}{2}\left(x_{\{k_1,k_2,q_1,q_2\}}^{2}-2j\frac{\pi fx_{\{k_1,k_2,q_1,q_2\}}}{\alpha_{\tau}B^{2}}
+\frac{\pi^{2}f^{2}}{(\alpha_{\tau}B^{2})^{2}}\right)},
\end{align}
\begin{align}
& X_{2}(f)=\frac{1}{B^{2}} \Bigg[\frac{1}{j2\pi x_{\{k_1,k_2,q_1,q_2\}}}\bigg(\bigg(e^{j\pi Bx_{\{k_1,k_2,q_1,q_2\}}} \nonumber \\
&-e^{j\pi (2f-B)x_{\{k_1,k_2,q_1,q_2\}}}\bigg) \mathbbm{1}_{\{0<f<B\}}
+ \bigg(e^{j\pi (2f+B)x_{\{k_1,k_2,q_1,q_2\}}} \nonumber \\
&-e^{-j\pi Bx_{\{k_1,k_2,q_1,q_2\}}}\bigg)\mathbbm{1}_{\{-B<f<0\}}\bigg)\Bigg]\mathbbm{1}_{\{x_{\{k_1,k_2,q_1,q_2\}}\neq 0\}} \nonumber \\
& + \hspace{-0.5mm} \frac{1}{B^{2}}\Big[(B \hspace{-0.5mm} - \hspace{-0.5mm} f)\mathbbm{1}_{\{0<f<B\}} \hspace{-0.5mm} + \hspace{-0.5mm} (B \hspace{-0.5mm} + \hspace{-0.5mm} f)\mathbbm{1}_{\{-B<f<0\}}\Big]\mathbbm{1}_{\{x_{\{k_1,k_2,q_1,q_2\}}=0\}},
\end{align}
 
Now, consider the case $x_{\{k_1,k_2,q_1,q_2\}} \neq 0$. 
The integral in (\ref{integral_3_2}) becomes
\begin{align}
&S_{\{k_1,k_2,q_1,q_2\}}^{(1)}=C_{3,1}(x_{\{k_1,k_2,q_1,q_2\}}) \nonumber \\
&\Bigg[e^{j\pi Bx_{\{k_1,k_2,q_1,q_2\}}}\underbrace{\int_{0}^{B}e^{-\frac{\pi^{2}f^{2}}{2\alpha_{\tau} B^{2}}-j\pi fx_{\{k_1,k_2,q_1,q_2\}}}df}_{\overset{\Delta}{=}S_{1,\{k_1,k_2,q_1,q_2\}}^{(1)}} \nonumber \\
& -e^{-j\pi Bx_{\{k_1,k_2,q_1,q_2\}}}\underbrace{\int_{0}^{B}e^{-\frac{\pi^{2}f^{2}}{2\alpha_{\tau} B^{2}}+j\pi fx_{\{k_1,k_2,q_1,q_2\}}}df}_{\overset{\Delta}{=}S_{2,\{k_1,k_2,q_1,q_2\}}^{(1)}} \nonumber \\
&+e^{j\pi Bx_{\{k_1,k_2,q_1,q_2\}}}\underbrace{\int_{-B}^{0}e^{-\frac{\pi^{2}f^{2}}{2\alpha_{\tau} B^{2}}+j\pi fx_{\{k_1,k_2,q_1,q_2\}}}df}_{\overset{\Delta}{=}S_{3,\{k_1,k_2,q_1,q_2\}}^{(1)}} \nonumber \\
& -e^{-j\pi Bx_{\{k_1,k_2,q_1,q_2\}}}\underbrace{\int_{-B}^{0}e^{-\frac{\pi^{2}f^{2}}{2\alpha_{\tau} B^{2}}-j\pi fx_{\{k_1,k_2,q_1,q_2\}}}df}_{\overset{\Delta}{=}S_{4,\{k_1,k_2,q_1,q_2\}}^{(1)}} \Bigg],
\label{integral_3_3}
\end{align}
where $C_{3,1}(x_{\{k_1,k_2,q_1,q_2\}})\overset{\Delta}{=}N_{0}\Omega_{\tau}^{2}\sqrt{\frac{\pi}{2\alpha_{\tau}}}\frac{e^{-\frac{\alpha_{\tau}B^{2}}{2}x_{\{k_1,k_2,q_1,q_2\}}^{2}}}{j2\pi x_{\{k_1,k_2,q_1,q_2\}}B^{2}}$, and the integrals in (\ref{integral_3_3}) are given by \\
$S_{1,\{k_1,k_2,q_1,q_2\}}^{(1)}=f_1\left(a_1, B,0,\pi x_{\{k_1,k_2,q_1,q_2\}}\right)$, \\
$S_{2,\{k_1,k_2,q_1,q_2\}}^{(1)}=f_1\left(a_1, B,0,-\pi x_{\{k_1,k_2,q_1,q_2\}}\right)$, \\
$S_{3,\{k_1,k_2,q_1,q_2\}}^{(1)}=f_1\left(a_1, 0,-B,-\pi x_{\{k_1,k_2,q_1,q_2\}}\right)$, \\
$S_{4,\{k_1,k_2,q_1,q_2\}}^{(1)}=f_1\left(a_1, 0,-B,\pi x_{\{k_1,k_2,q_1,q_2\}}\right)$. \\  

Now, consider the case $x_{\{k_1,k_2,q_1,q_2\}}=0$.
The integral in (\ref{integral_3_2}) becomes
\begin{eqnarray}
S^{(2)} & \hspace{-2mm} = & \hspace{-2mm} C_{3,2}\Bigg[B\underbrace{\int_{0}^{B}e^{-\frac{\pi^{2}f^{2}}{2\alpha_{\tau} B^{2}}}df}_{\overset{\Delta}{=}S_1^{(2)}}-\underbrace{\int_{0}^{B}fe^{-\frac{\pi^{2}f^{2}}{2\alpha_{\tau} B^{2}}}df}_{\overset{\Delta}{=}S_2^{(2)}} \nonumber \\
& \hspace{-2mm} & \hspace{-2mm} +B\underbrace{\int_{-B}^{0}e^{-\frac{\pi^{2}f^{2}}{2\alpha_{\tau} B^{2}}}df}_{\overset{\Delta}{=}S_3^{(2)}} + \underbrace{\int_{-B}^{0}fe^{-\frac{\pi^{2}f^{2}}{2\alpha_{\tau} B^{2}}}df}_{\overset{\Delta}{=}S_4^{(2)}} \Bigg],
\label{integral_3_4}
\end{eqnarray}
where the constant $C_{3,2}=N_{0}\Omega_{\tau}^{2}\sqrt{\frac{\pi}{2\alpha_{\tau}}}\left(\frac{1}{B^{2}}\right)$, and the integrals in (\ref{integral_3_4}) are obtained as
$S_1^{(2)}=\frac{\sqrt{\pi}}{2\sqrt{a_1}}\mathrm{erf}\big(\sqrt{a_1}B\big)$,
$S_2^{(2)}=\frac{1}{2a_1}\big(1-e^{-a_1 B^{2}}\big)$,
$S_3^{(2)}=\frac{\sqrt{\pi}}{2\sqrt{a_1}}\mathrm{erf}\big(\sqrt{a_1}B\big)$,
and
$S_4^{(2)}=\frac{1}{2a_1}\big(e^{-a_1 B^{2}}-1\big)$.

The erf(.) functions in the derived expressions can be computed using accurate closed-form approximations for the erf(.) function \cite{erf1},\cite{erf2}.

\end{document}